\documentclass[11pt,preprint]{aastex}

\newcommand{\teff}{$T_{\rm eff}$}
\newcommand{\vi}{$V - I$}
\newcommand{\LL}{${\lambda}{\lambda}$}
\newcommand{\lgf}{log $gf$}
\newcommand{\spr}{$s$-process}

\newcommand{\cratio}{$^{12}$C/$^{13}$C ratio}

\received{28 February 2000}
\accepted{17 May 2000}

\slugcomment{To appear in {\em The Astrophysical Journal}, September 2000}

\shortauthors{Cavallo \& Nagar}
\shorttitle{Abundances in M3 and M13 Giants}

\begin{document}

\title{Aluminum Abundances, Deep Mixing and the Blue-Tail Second-Parameter
 Effect in the Globular Clusters M~3 and M~13}

\author{Robert M. Cavallo\altaffilmark{1,2}}
\affil{Laboratory for Astronomy and Solar Physics, NASA/Goddard Space Flight
 Center, Greenbelt, MD 20771}
\email{rob@shemesh.gsfc.nasa.gov}

\and

\author{Neil M. Nagar\altaffilmark{1}}
\affil{Department of Astronomy, University of Maryland, College Park, MD 20742}

\altaffiltext{1}{Visiting Astronomer, Kitt Peak National Observatory. 
KPNO is operated by AURA, Inc.\ under contract to the National Science
Foundation.} 

\altaffiltext{2}{NRC Research Associate}

\begin{abstract}
\begin{small}
We analyze high resolution, high signal-to-noise spectra
 of six red-giant-branch (RGB) stars in the globular cluster M~3
 (NGC~5272) and three in M~13 (NGC~6205) that were obtained with the
 Mayall 4-meter telescope and echelle spectrometer on Kitt Peak.
The spectra include lines of O, Na, Mg, Al, Si, Ca, Ti, V, Mn, Fe and Ni.
We also analyze the [Al/Fe] values of 96 RGB stars in M~13 covering
 the brightest 3.5 magnitudes, which include 66 measurements that were
 derived from moderate resolution, low signal-to-noise spectra 
 obtained with the WIYN 3.5-meter telescope and Hydra multi-object
 spectrograph, also on Kitt Peak.
In addition, we compile from the literature and inspect the [Na/Fe] values
 of 119 RGB stars in M~13.
We test for bimodality in the [Al/Fe] and [Na/Fe] distributions using the KMM
 algorithm and find that the [Al/Fe] values in M~13 are distributed
 bimodally at all points along the RGB that were observed,
 while the [Na/Fe] values are bimodal only over the brightest two magnitudes.
The ratios of Al-enhanced to Al-normal and Na-enhanced to Na-normal giants
 increase towards the tip of the RGB in M~13, which is suggestive of
 deep mixing in this cluster.
The limited M~3 data exhibit a bimodal distribution of [Al/Fe] values 
 and are suggestive of no deep mixing; however, they are too few to be 
 conclusive.
We further test for a relationship between deep mixing on the RGB and a
 second parameter that can create the extended blue tail seen along
 the  horizontal-branches of some clusters by using an ``instantaneous''
 mixing algorithm, which we develop here.
We conclude that the data for both clusters are consistent with
 deep mixing as a ``blue-tail second parameter'', and we suggest future
 observations to further constrain the results.
Finally, we offer a solution to the problem of over producing sodium during
 deep mixing that is based on the depletion of $^{22}$Ne in
 asymptotic-giant-branch stars and suggest that pollution might best be
 traced by {\spr} elements in the Sr-Y-Zr peak.
\end{small}
\end{abstract}

\keywords{globular clusters: individual (M~3,M~13)
 --- stars: abundances --- stars: horizontal branch --- stars: late-type
 --- stars: Population II}

\section{INTRODUCTION}

According to canonical stellar evolution models, the by-products of the
 nuclear processing around the hydrogen-burning shell (H~shell)
 of low-mass red-giant-branch (RGB) stars should remain confined to the stellar
 interior; however, observations over the past 25 years have shown
 star-to-star variations in the elements C, N, O, Na, Mg and Al, among
 others, on the surfaces of globular cluster red giants (see Kraft 1994,
 Briley et al. 1994 and Cavallo 1998a for detailed reviews of the observations).
In particular, the data show evidence of the CNO cycle that dominates the
 energy production in such stars: C and O are anticorrelated with N, while
 the {\cratio} is near the equilibrium value of 4 in many clusters
 \citep{SS91,S96b,BSKL97,BSSBN97,ZWB96}.
While the first dredge-up phenomenon \citep{Iben67} does alter the carbon
 and nitrogen abundances slightly, it cannot account for the observed large
 variations of these elements and their isotopic ratios, nor can it account
 for the variations of the other elements.
In addition, some elements show evidence for gradual changes along
 the RGB, indicating that something is occurring during the course of
 evolution to facilitate these alterations.
For example, C becomes more depleted with decreasing $V$ in
 the clusters M~15, M~55, M~92 and NGC~6397 \citep{BDG79,Carbon1982,TCLSK83,
 BBHD90}.

Two separate approaches have been developed to address the observations.
One assumes that some form of non-canonical mixing occurs along the
 RGB, which gradually brings material from around the H~shell to the stellar
 surface \citep[hereafter, SM79]{SM79}.
Models by SM79, \citet{DD90}, \citet{LHS93}, \citet{CSB96}, \citet{DW96} and 
 \citet[hereafter, CSB98]{CSB98} have shown that most variations 
 along the brighter part of the RGB can be explained by nuclear processing
 around the H~shell combined with mixing.
The source of mixing is generally assumed to be rotationally induced
 meridional circulation currents (SM79); although, other theories abound
 \citep{LHZ97,FAK99}.
The observations by \citet{Peterson83} that show the horizontal-branch
 (HB) stars in M~13, a cluster with large variations of oxygen and
 aluminum on the RGB, rotating nearly a factor of two faster than the HB stars
 in M~3, a cluster with a composition similar to M~13, but with
 less extreme abundance variations along it's RGB, support the SM79
 hypothesis.

The second approach assumes that some of the variations,
 particularly those of the heavier elements, are primordial in
 nature, perhaps originating in the processed envelopes
 of intermediate-mass asymptotic-giant-branch (AGB) stars that were
 shed into the nascent cluster environment \citep{CD81}.
While it has been shown that this scenario cannot account for all the
 variations \citep{DWW97}, some aspects of it are plausible in light of
 the data.
For example, observations of CN-band strength and sodium variations on the
 upper main sequence of 47~Tuc, sodium enhancements on the subgiant branch 
 of M~92 and enhancements in the neutron-capture elements in some clusters
 all point to primordial origins
 \citep{BBSH89,BHB91,S96a,BSSLBH96,CCBHS98,KSB98,Ivans99}.
The most likely solution to the abundance anomaly problem probably
 involves a combination of both scenarios, where primordial pollution
 is present in the cluster, but mixing later plays a role in adjusting 
 the abundance patterns (see, e.g., Denissenkov et al. 1998 and Briley et al.
 1999), an approach we examine here.

This paper focuses on determining the chemical abundances
 in the red giants of the globular clusters M~3 (NGC~5272) and M~13
 (NGC~6205) from high resolution,
 high signal-to-noise echelle spectra obtained with the Mayall 4-meter
 telescope on Kitt Peak.
We choose these two clusters because they are often considered a classical
 ``second-parameter'' pair since they have markedly different HB's,
 despite having similar [Fe/H]\footnote{
 We use the usual notation: [X/Y]~=log(X/Y)$_{\star}$~$-$~log(X/Y)$_{\sun}$.}
 values, the first parameter.
We discuss the hypothesis that deep mixing along the RGB, which we define 
 as mixing that penetrates the H~shell, brings helium to the surface and
 affects the HB morphology as a second parameter that creates the extended
 blue tail in M~13 \citep{AVS97a,AVS97b}.
One oft-quoted choice for the second parameter is a relative age difference
 between M~3 and M~13 \citep{FPB97,SCD97,Chaboyer98}; however, this is not
 borne out by the photometry \citep{JB98,vdB99,FG99} and leaves open the
 need for a qualified alternative.

While M~13 is by far the most well-studied cluster for abundance variations,
 the data for M~3 are lacking.
One goal of this paper is to increase our knowledge of the chemical 
 abundances in this latter cluster so that it can be compared with M~13 in
 greater detail.

The outline of this paper is as follows:
 we describe the observations, data reduction approach, abundance analysis
 technique and abundance results in sections 1, 2, 3 and 4, respectively.
In section 5 we discuss evidence for and the implications of deep mixing
 along the RGB, and we give our conclusions in section 6.
In the appendix we derive the instantaneous mixing algorithm that is used 
 in section 5.

\section{OBSERVATIONS}

We chose five bright giants in M~3 (MB~4, vZ~205, vZ~297, vZ~1000 and vZ~1127)
 and three in M~13 (L~262, L~324 and L~414) from previous studies by the
 ``Lick/Texas'' group \citep{KSLP92,KSLS93,KSLSB95} that show
 evidence for abundance variations but have no information regarding aluminum.
The observations were made on the nights of 29-31 May 1998 using the
 echelle spectrograph on the Mayall 4-meter telescope.
The echelle setup used - echelle grating 31.6-63{\arcdeg}, 
 cross grating 226-1, long-focus camera and the T2KB CCD - 
 resulted in continuous spectral coverage between 5500~{\AA}
 and 8800~{\AA} at a dispersion of 0.08 {\AA} per CCD pixel,
 i.e. resolution, R$\sim$ 30,000 at 6000 {\AA} over a 2.5 pixel resolution
 element. 
The central wavelength on the CCD was chosen to maximize the signal-to-nose 
 ratio near the {\LL}6696, 6698 {\AA} Al I lines. 
The seeing was typically between 1{\arcsec} and 1{\farcs}2 on all three nights,
 causing us to use a slit-width of 1\farcs5, and a slit length of 7\farcs5.
Each 4 ${\times}$ 30 min. exposure of a target star was sandwiched between
 two 15 s exposures of a ThAr comparison lamp, which was observed
 at the same telescope position and slit position angle (P.A.).
To facilitate subtraction of telluric lines we observed one or two fast
 rotating B stars each night at various airmasses.

In addition to our data, Dr. M. Briley provided us with the reduced spectrum
 of the M~3 giant AA and a fast rotating B star, taken on 21 March 1998 
 with the McDonald Observatory 2.7-m telescope and 2D-Coude echelle
 spectrograph.
His set-up yielded an non-continuous wavelength coverage from 4070 {\AA} to
 10,500 {\AA} with $R ~{\sim} ~$ 50,000 over a resolution element.

A log of all the observations with estimates of the signal-to-noise ratio
 around the {\LL}6696, 6698 {\AA} Al I region is given in Table 1.
The photometry is from \citet{Ferr97} for M~3 and from \citet{CM79} for M~13.
The locations of the program stars in their respective color-magnitude diagrams
 are given in Figure 1 for M~3 and Figure 2 for M~13.
In the following discussions, we refer to each star by its most commonly
 used identification, which in some cases is its alternate name.

\begin{figure}
\epsscale{0.50}
\plotone{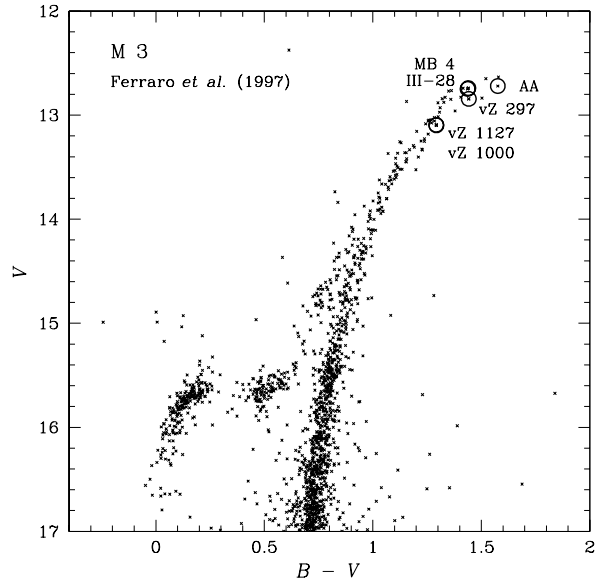}
\caption[Cavallo.fig01.eps]{ Color-magnitude diagram for M~3,
 with the program stars circled and labeled.
 The data are from \citet{Ferr97}.}
\end{figure}
\begin{figure}
\plotone{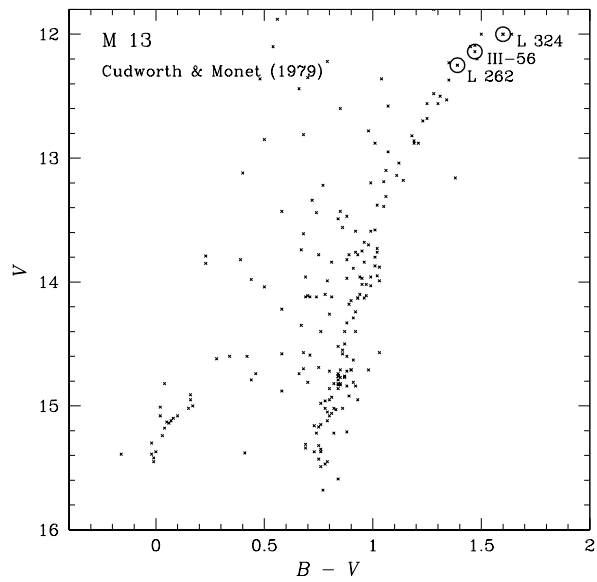}
\caption[Cavallo.fig02.eps]{As Figure 1, except for M~13 with the data from
 \citet{CM79}.}
\end{figure}

\section{CCD PROCESSING AND SPECTRA EXTRACTION}

The data were reduced using standard IRAF\footnote{IRAF is distributed by
 the National Optical Astronomy Observatories, which are operated by the
 Association of Universities for Research in Astronomy, Inc., under
 cooperative agreement with the National Science Foundation.}\citep{IRAF}
 tasks (version 2.11.1), following the reduction procedure outlined in ``Users
 guide to reducing Echelle spectra with IRAF'', by D. Wilmarth \& J.
 Barnes\footnote{available by anonymous ftp to \url{ftp://iraf.noao.edu/iraf/docs}}. 
Zero frames were taken on each night, but since $>$ 99.9\% of the pixels 
 had zero values within the 5 e$^-$ r.m.s. noise of the T2KB CCD, we only
 zero-corrected ``hot'' (zero values $>$ 5 e$^-$) pixels.
Quartz flat exposures were taken on each of the three nights and used
 to flat field the target stars and to determine ``dead'' pixels.

After initial processing of the CCD target star data, we used the IRAF task
 APSCATTER to correct for scattered light between the orders.
The orders were then extracted to single dimensional spectra using the
 task APSUM with variance weighting, and then wavelength calibrated using
 the closest (in time) ThAr spectrum, taken at the same telescope position
 and slit P.A. as the target star.
Orders that contained telluric lines were corrected using the task
 TELLURIC and a comparison spectrum of a fast-rotating B star that
 was observed at an airmass similar to the program star.
Finally, each spectrum was shifted into the rest frame and flattened
 by fitting a spline through the continuum.

The spectrum for the star M~3 AA was given to us in its extracted form
 and required only correction for telluric lines in some orders.

Figure 3 shows the spectra around the Al~I region for all the M~3 and
 M~13 giants.
The variation in the Al~I line strengths is quite apparent from one star to
 the next.

\begin{deluxetable}{lllllll}
\tabletypesize{\footnotesize}
\tablewidth{0pc}
\tablenum{1}
\tablecaption{Observing Log}
\tablecolumns{7}
\tablehead{
 \colhead{Star }          & 
 \colhead{Alt.}           & 
 \colhead{$V$}            & 
 \colhead{{\bv}}          & 
 \colhead{Date Obs.}      &
 \colhead{Exposure}       & 
 \colhead{S/N}            \\ 
 \multicolumn{2}{c}{}     & 
 \colhead{mag.}           &
 \colhead{mag.}           &
 \colhead{U.T.}           &
 \colhead{min.}           &
 \colhead{est.}
}

\startdata
\cutinhead{M~3}
vZ~238  & AA, SK 586, F 12959 & 12.72  & 1.58 & 21-23 March 1998{\tablenotemark{a}} & 180 & 110 \\
vZ~752  & MB 4, F 14194 & 12.74  & 1.44 & 1 June 1998                         & 120 & 115 \\
vZ~205  & III-28, SK 617, F 16682  & 12.75  & 1.44 & 31 May 1998                         & 120 & 95  \\
vZ~297  & SK 525, F 9  & 12.84  & 1.44 & 30 May 1998                         & 120 & 105 \\
vZ~1127 & MB 26, F 14246 & 13.09  & 1.29 & 31 May/1 June 1998                  & 180 & 135 \\
vZ~1000 & SK 297, F 17307 & 13.10  & 1.29 & 31 May/1 June 1998                  & 180 & 90  \\
\\
\cutinhead{M~13}
L~324   & V 11, CM~425 & 12.00  & 1.60 & 1 June 1998                         & 120 & 160 \\
L~414   & III-56, CM~176 & 12.14  & 1.47 & 30 May 1998                         & 120 & 140 \\
L~262   & CM~476  & 12.25  & 1.39 & 31 May 1998                         & 120 & 140 \\
\enddata
\tablenotetext{a}{Observed by M. Briley}
\tablecomments{Star names taken from the catalogs of \citet[vZ]{vZ08} and
 \citet[L]{L05}. Alternate names are from \citet[AA, III-28]{San53},
 \citet[III-56]{Arp55}, \citet[V11]{Hogg73}, \citet[CM]{CM79}, \citet[SK]{SK82},
 \citet[F]{Ferr97} and Michael Bolte (1998, priv comm. MB).}
\end{deluxetable}

\begin{figure}
\plotone{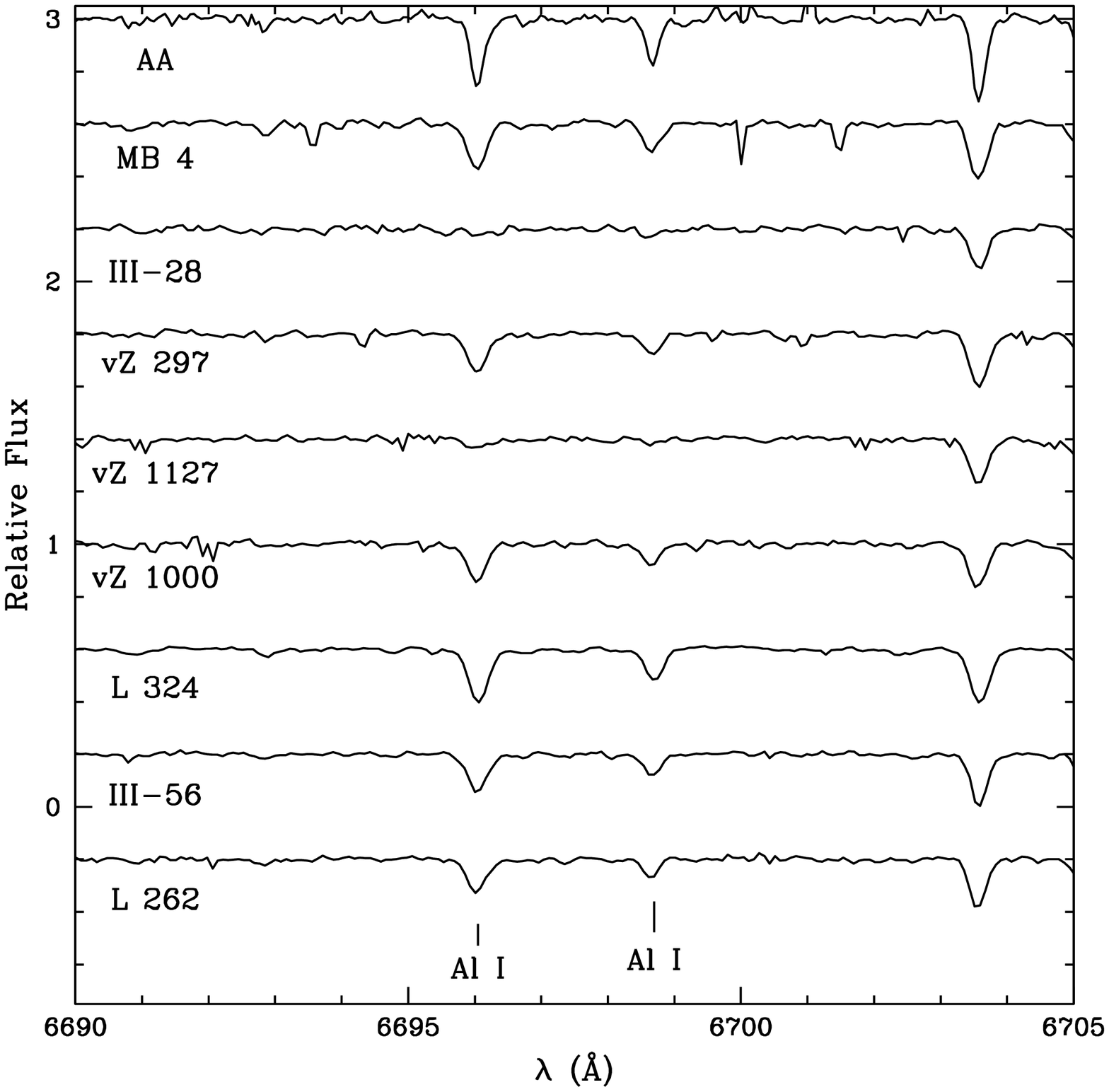}
\caption[Cavallo.fig03.eps]{The Al I spectral region for all M~3 and M~13
 stars observed.
 The ordinate is in relative units, with each spectrum offset from
 the others. 
 The Al I lines at ${\lambda\lambda}6696,6698$~{\AA} are labeled.}
\end{figure}

\section{ABUNDANCE ANALYSIS}
\subsection{Equivalent Widths, Line Lists and Line Parameters}
The free spectral range for the Mayall 4-meter data is less than the width
 of the CCD so that each line appears at least twice in our spectra,
 with one line usually towards the center of the chip where the noise 
 is minimized.
Unfortunately, combining adjacent orders to increase the number of
 photon counts resulted in a lowered signal-to-noise ratio since the
 edges of the chip contained much lower quality spectra than the center.
We measured the equivalent widths of the line closer to the center of the
 CCD and present the results in Table 2 (found after {\bf REFERENCES}).
The equivalent widths were determined by summing the flux in a line if
 the line was cleanly separated from other lines or by fitting a Gaussian
 in more crowded regions.
Some data are excluded from the table for one of several reasons:
 1) In the case of M~3 AA, some lines fell between the orders;
 2) the line was too weak to be measured (our minimum measurable equivalent
    width was about 5 m{\AA}, depending on the signal-to-noise ratio);
 3) the line was contaminated by either a bad column or a cosmic ray hit; or
 4) in very few cases, the line gave abundance results that were anomalously
    discordant with the mean and r.m.s. deviation of the rest of the lines 
    of the same species for unknown reasons and was rejected.

\subsubsection{Fe lines}

The iron lines were chosen from Kurucz's CD-ROM~\# 23\footnote{
 \url{http://cfa-www.harvard.edu/amdata/ampdata/amdata.html}}
 and the solar atlas of \citet{MMH66}.
They were used to determine both the iron abundances and the model-atmosphere
 parameters and were thus selected to have a broad range in excitation 
 potentials and oscillator strengths, which were adopted without modification
 from the empirically derived tables of \citet{Nav94} and \citet{Bie91} for
 the Fe~I and Fe~II lines, respectively.
We rejected Fe I lines that were listed by \citet{Nav94} as showing blends with
 other iron lines or uncertainties in their energy levels of more than
 0.005 cm$^{-1}$.
For the sake of comparison, we measured the equivalent widths of our
 chosen iron lines in the visible and near-infrared solar
 spectra\footnote{NSO/Kitt Peak FTS data used here were produced by NSF/NOAO.}
 of \citet{WHL93,WHL98}\footnote{
 available at \url{http://www.nso.noao.edu/diglib/ftp.html}}
  and folded them through the \citet{HM74} solar model atmosphere with a
  microturbulent velocity of 1.0~km s$^{-1}$.
We adjusted the {\lgf}'s to reproduce an assumed solar iron abundance of
 log~${\epsilon}({\rm Fe})_\odot ~= ~7.52$\footnote{log~${\epsilon}$(X) =
 log N(X/H) + 12.00, where N is the number abundance. For an informative
 debate about the preferred solar iron abundance we refer the reader to
 the papers of \citet{BLS95}, \citet{HKB95}, \citet{BSL95} and \citet{KSR96}.
 For consistency with the Lick/Texas studies we adopt the 7.52 value.}.
This results in an average difference between the two sets of oscillator
 strengths (in the sense of solar model $-$ laboratory) of $+0.09 ~{\pm} 
~0.12$ for Fe I and $+0.05 ~{\pm} ~0.08$ for Fe II (see Table 3).
Both differences show a slight trend toward lower oscillator strengths from
 the laboratory measurements ($< ~1 ~{\sigma}$), which in
 turn would cause the determined iron abundances to be overestimated by 0.09 
 and 0.05 dex from the Fe I and Fe II lines, respectively.

\subsubsection{EW comparisons}

In the main panel of Figure 4 we compare the equivalent widths of our
 iron lines with those from the earlier Lick/Texas studies.
The solid line in the figure has a 45 degree slope and represents perfect
 agreement between the two data sets.
The Lick/Texas data tend toward higher equivalent widths relative
 to our data, especially above ${\sim}$100 m{\AA}.
The average differences between the two data sets (in the sense of 
 Lick/Texas $-$ present work) are 6.9 ${\pm}$ 9.7 m{\AA} for Fe I and 5.9
 ${\pm}$ 9.7 m{\AA} for Fe II.
We attribute the differences to several factors.
First, our present data have higher signal-to-noise ratios
 than the Lick/Texas data (90 to 160 compared with 40 - 100), which reduces the
 level of uncertainty in placing the continuum.
Second, the Lick/Texas spectra have higher resolution, $R ~{\sim}
 ~48,000$ compared with $R ~{\sim} ~30,000$, which helps separate lines
 from the continuum and other lines.
Third, the two datasets were reduced using two separate software packages
 that apply scattered-light corrections differently (see, e.g., Sneden et al.
 1991), which can affect the continuum levels and depths of each line.
Fourth, measuring equivalent widths is subjective and two different 
 observers can get different results from the same data.
For example, the inset in Figure 4 shows the equivalent widths for two
 separate observations of the M~13 giant III-56 by the Lick/Texas group and
 demonstrates how much variability can be present in the data even
 with consistent reduction techniques (see also Kraft et al. 1993, Figure 1).
Fifth, the methods of measuring equivalent widths differ: we use both
 a Simpson's rule technique (i.e., direct integration) and Gaussian fits,
 while the Lick/Texas results come from Gaussian fits for their earlier
 papers and both techniques for \citet{Kraft97}.

\begin{figure}
\plotone{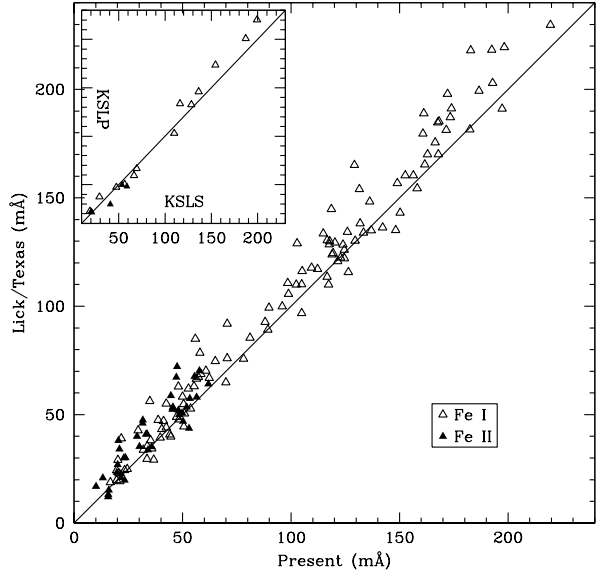}
\caption[Cavallo.fig04.eps]{Main panel: Our Fe I and Fe II equivalent widths
 compared with the measurements from the Lick/Texas group.
 Inset: Comparison of Fe I and Fe II equivalent width measurements for the
 M~13 giant III-56 from two different studies by the Lick/Texas group
 [Kraft et al. 1992 (KSLP) and Kraft et al. 1993 (KSLP)].
 The scale on the ordinate is identical to the abscissa.
 The solid line in both plots has a slope of 1. }
\end{figure}

\subsubsection{Other lines}

We determined the oscillator strengths for the other elements in our
 study by measuring the equivalent widths in the solar spectrum 
 and adjusting the {\lgf} values until we derived the \citet{AG89}
 solar abundances using the \citet{HM74} solar model.
Table 3 lists the average differences between the values we determined
 and the literature values, which are mostly from \citet{Thev90}, who
 did a similar differential analysis using an older solar spectrum with
 a MARCS \citep{MARCS} solar model and a microturbulent velocity of 0.6 km
 s$^{-1}$.
With the exception of the Mn I lines, we used our derived {\lgf}'s to derive
 the elemental abundances.
The two Mn I lines in the solar spectrum (we discard a third line at
 ${\lambda}$6022 {\AA} since it blends with a nearby Fe I feature) suffer 
 from hyperfine splitting effects.
To avoid detailed calculations, we adopt the recommended {\lgf}'s 
 from \citet{Thev90} for these two manganese lines.

\begin{deluxetable}{lllcc}
\tabletypesize{\small}
\tablenum{3}
\tablewidth{0pc}
\tablecaption{Comparison of Oscillator Strengths}
\tablecolumns{5}
\tablehead{
 \colhead{Species}         &
 \colhead{Avg. Diff}       &
 \colhead{${\pm}{\sigma}$} &
 \colhead{\# Lines}        &
 \colhead{Ref.} 
}
\startdata
Na I  & $+0.015$ & 0.021 & 2  & T90     \\
Mg I  & $+0.263$ & 0.349 & 3  & T90, K  \\
Al I  & $+0.035$ & 0.052 & 6  & T90, K  \\
Si I  & $+0.059$ & 0.108 & 9  & T90, K  \\
Ca I  & $+0.203$ & 0.059 & 8  & T90, K  \\
Ti I  & $-0.025$ & 0.113 & 8  & T90, K  \\
Ti II & $-0.120$ & 0.282 & 2  & T90     \\
V I   & $+0.048$ & 0.084 & 18 & T89, T90\\
Mn I  & $+0.615$\tablenotemark{a} & 0.064\tablenotemark{a} & 2  & T90 \\
Ni I  & $+0.203$ & 0.097 & 12 & T90 \\
Fe I  & $+0.088$ & 0.123 & 24 & N \\
Fe II & $+0.048$ & 0.077 & 8  & B\\
\enddata
\tablenotetext{a}{Used values from \citet{Thev90}.}
\tablerefs{
 B = \citet{Bie91}; K = Kurucz CD-ROM~\#23; N = \citet{Nav94};
 T89 = \citet{Thev89}; T90 = \citet{Thev90}.
 }
\end{deluxetable}

\subsection{Model Parameters: Effective Temperature, Gravity and
            Microturbulence}
\subsubsection{Spectroscopic models}

We constructed models using the Fe I and Fe II lines with the MARCS model
 atmosphere code and the MOOG abundance-analysis code \citep{MOOG}.
The initial models were built with the parameters determined by the
 Lick/Texas group and were constructed with the alpha elements enhanced 
 by 0.4 dex in accordance with previous observations of cluster giants 
 (see, e.g., Kraft et al. 1997).
We iteratively ran MOOG and MARCS to refine the models until the
 derived abundances were independent of excitation potential, line width
 and ionization level.
We checked our final choice of model parameters by independently
 using the Ni~I lines to determine {\teff}, the Ni~I and Ti~I lines to 
 determine the microturbulent velocity, $v_t$, and the Ti~I and Ti~II lines
 to determine log~$g$.
The results were generally in agreement with the more numerous iron lines
 and allowed us to estimate systematic errors in the model parameters
 determined from spectroscopy: ${\Delta}T_{\rm eff} ~{\sim} ~{\pm}30 ~{\rm K},
 ~{\Delta}{\rm log}(g) ~{\sim} ~{\pm}0.2 ~{\rm cm~s^{-2}}, ~{\rm and} 
 ~{\Delta}v_{t} ~{\sim} ~{\pm}0.15 {\rm km~s^{-1}}$.

Our final spectroscopically determined model parameters are given in Table 4,
 along with the original Lick/Texas model parameters, and are the parameters
 we used to derive the elemental abundances.
The effective temperatures generally agree to less than 100 K, while 
 our gravities are typically lower than the Lick/Texas values by 0.15 
 ${\pm}$ 0.27 dex and our microturbulent velocities are lower
 by 0.09 ${\pm}$ 0.11 km s$^{-1}$.
The differences are consistent with the error estimates derived from the
 nickel and titanium lines and are not surprising given the differences in
 the equivalent widths and the choices of lines and line parameters
 between the two studies.

\begin{deluxetable}{lccccccc}
\tablewidth{0pc}
\tablenum{4}
\tablecolumns{8}
\tablecaption{Spectroscopic Models}
\tablehead{
                       \colhead{} & 
 \multicolumn{3}{c}{Present Work} & 
 \colhead{}                       &
 \multicolumn{3}{c}{Lick/Texas}   \\
 \cline{2-4} \cline{6-8}
 \colhead{Star}    &
 \colhead{\teff}   &
 \colhead{log $g$} &
 \colhead{$v_t$}   &
 \colhead{}        &
 \colhead{\teff}   &
 \colhead{log $g$} &
 \colhead{$v_t$}
}
\startdata
\cutinhead{M~3}
AA      & 4050 & 0.40 & 2.27 && 4000 & 0.40 & 2.25 \\
MB 4    & 4060 & 0.45 & 2.05 && 3925 & 0.30 & 2.15 \\
III-28  & 4175 & 0.55 & 1.84 && 4160 & 0.75 & 1.75 \\
vZ~297  & 4050 & 0.25 & 1.98 && 4070 & 0.70 & 2.25 \\
vZ~1127 & 4300 & 1.00 & 1.98 && 4225 & 0.90 & 2.00 \\
vZ~1000 & 4200 & 0.65 & 1.94 && 4175 & 0.45 & 2.10 \\
\cutinhead{M~13}
L~324   & 3990 & 0.10 & 2.34 && 4050 & 0.50 & 2.50 \\
III-56  & 4030 & 0.20 & 2.13 && 4100 & 0.65 & 2.25 \\
L~262   & 4160 & 0.50 & 1.89 && 4180 & 0.80 & 2.00 \\
\enddata
\end{deluxetable}

\subsubsection{Photometric models}

We used recent photometry of our cluster giants, given in Tables 5a and 5b,
 to derive alternative model atmosphere parameters.
The {\bv} and {\vi} data were used to derive effective temperatures
 based on a 12 Gyr isochrone that was constructed with [Fe/H] = $-1.54$
 and [${\alpha}$/Fe] = $+0.3$ dex.
The luminosities and {\teff}'s for the models in the isochrone were computed
 by Dr. D.  VandenBerg while Dr. R. Bell performed the luminosity-M$_V$ 
 and the color-temperature transformations.
The age of the isochrone (mass) fixes the gravities, while the microturbulent
 velocities are still determined from spectroscopy.
The ${V - K}$ calibrations are from \citet{CFP78}.
We present the results of the photometric calibrations in Tables 6a and
 6b.
Using the extrema from the photometric and spectroscopic parameters we
 derived new model atmospheres from which we determined a range in the
 abundances allowed by the uncertainties in the models.

\begin{deluxetable}{lcccccccccccccccccc}
\tabletypesize{\footnotesize}
\rotate
\tablewidth{0pc}
\tablenum{5a}
\tablecaption{Photometry for M 3 Giants}
\tablehead{
                                            \colhead{} & 
 \multicolumn{3}{c}{Ferraro {\em et al.}} & \colhead{} & 
 \colhead{Rood}                           & \colhead{} &
 \multicolumn{2}{c}{MB}                   & \colhead{} & 
 \colhead{vB}                             & \colhead{} & 
 \multicolumn{2}{c}{SK}                   & \colhead{} & 
 \multicolumn{2}{c}{Cudworth}             & \colhead{} &
 \colhead{CFP}                            \\
 \cline{2-4} \cline{6-6} \cline{8-9} \cline{11-11} \cline{13-14} \cline{16-17} \cline{19-19} \\
 \colhead{Star}    &
 \colhead{$V$}     &
 \colhead{\bv}     &
 \colhead{$V - I$} & \colhead{}        & \colhead{$V - I$} & \colhead{}        & \colhead{$V$}     & \colhead{\bv}     & \colhead{}        & \colhead{$V - I$} & \colhead{}        & \colhead{$V$}     & \colhead{\bv}     & \colhead{}        & \colhead{$V$}     & \colhead{\bv}     & \colhead{}        & \colhead{$V - K$} } 
\startdata
AA      & 12.72 & 1.58 & \nodata && 1.46 && \nodata & \nodata && \nodata && 12.71   & 1.56    && 12.69   & 1.57    && 3.43    \\
MB 4    & 12.74 & 1.44 & 1.47    && 1.45 && 12.75   & 1.64    && \nodata && \nodata & \nodata && \nodata & \nodata && \nodata \\
III-28  & 12.75 & 1.44 & \nodata && 1.45 && \nodata & \nodata && 1.32    && 12.80   & 1.37    && 12.81   & 1.37    && 3.21    \\
vZ 297  & 12.84 & 1.44 & \nodata && 1.42 && \nodata & \nodata && \nodata && 12.85   & 1.43    && 12.89   & 1.42    && \nodata \\
vZ 1127 & 13.09 & 1.29 & 1.47    && 1.35 && \nodata & \nodata && \nodata && \nodata & \nodata && 12.93   & 1.23    && \nodata \\
vZ 1000 & 13.10 & 1.29 & 1.39    && 1.35 && \nodata & \nodata && \nodata && 13.03   & 1.29    && 13.01   & 1.40    && \nodata \\
\enddata
\end{deluxetable}

\begin{deluxetable}{lccccc}
\tablewidth{0pc}
\tablenum{5b}
\tablecaption{Photometry for M 13 Giants}
\tablehead{
                            \colhead{} & 
 \multicolumn{2}{c}{CM79} & \colhead{} &
 \colhead{CFP}              \\
 \cline{2-3} \cline{5-5}    \\
 \colhead{Star}    &
 \colhead{$V$}     &
 \colhead{\bv}     &
 \colhead{}        &
 \colhead{$V - K$}
}
\startdata
L 324  & 12.00 & 1.60 && \nodata \\
III-56 & 12.14 & 1.47 && 3.31    \\
L 262  & 12.25 & 1.39 && \nodata \\
\enddata
\end{deluxetable}

\begin{deluxetable}{lcccccccccccccc}
\tabletypesize{\footnotesize}
\tablewidth{0pc}
\tablenum{6a}
\tablecaption{Photometric Models for M 3 Giants}
\tablehead{
                            \colhead{} & 
 \multicolumn{2}{c}{$V - I_{\rm F97}$} & 
 \colhead{}                            &
 \multicolumn{2}{c}{\bv$_{\rm F97}$}   & 
 \colhead{}                            &
 \multicolumn{2}{c}{$V - I_{\rm FR}$}  & 
 \colhead{}                            &
 \multicolumn{2}{c}{$V - I_{\rm vB}$}  & 
 \colhead{}                            &
 \multicolumn{2}{c}{$V - K_{\rm CFP}$} \\
 \cline{2-3} \cline{5-6} \cline{8-9} \cline{11-12} \cline{14-15}\\
 \colhead{Star}          &
 \colhead{$T_{\rm eff}$} &
 \colhead{log $g$}       &
 \colhead{}              &
 \colhead{$T_{\rm eff}$} &
 \colhead{log $g$}       &
 \colhead{}              &
 \colhead{$T_{\rm eff}$} &
 \colhead{log $g$}       &
 \colhead{}              &
 \colhead{$T_{\rm eff}$} &
 \colhead{log $g$}       &
 \colhead{}              &
 \colhead{$T_{\rm eff}$} &
 \colhead{log $g$}
}
\startdata
AA      & \nodata & \nodata && 3950 & 0.31 && 4061 & 0.57 && \nodata & \nodata && 4000    & 0.70    \\
MB 4    & 4060    &  0.57   && 4046 & 0.54 && 4067 & 0.58 && \nodata & \nodata && \nodata & \nodata \\
III-28  & \nodata & \nodata && 4048 & 0.54 && 4070 & 0.59 && 4232    & 0.90    && 4100    & 0.80    \\
vZ 297  & \nodata & \nodata && 4043 & 0.54 && 4102 & 0.65 && \nodata & \nodata && \nodata & \nodata \\
vZ 1127 & 4059    &  0.56   && 4182 & 0.80 && 4189 & 0.81 && \nodata & \nodata && \nodata & \nodata \\
vZ 1000 & 4088    &  0.62   && 4182 & 0.80 && 4190 & 0.81 && \nodata & \nodata && \nodata & \nodata \\
\enddata
\end{deluxetable}

\begin{deluxetable}{lcc}
\tablewidth{0pc}
\tablenum{6b}
\tablecaption{Photometric Models for M 13 Giants}
\tablehead{
 \colhead{Star}          &
 \colhead{$T_{\rm eff}$} &
 \colhead{log $g$}       \\
}
\startdata
L 324  & 3905 & 0.27 \\
III-56 & 4020 & 0.49 \\
L 262  & 4089 & 0.62 \\
\enddata
\end{deluxetable}

\subsection{Results}

Tables 7a, 7b and 7c present the final results of our abundance analysis
 for the iron-peak elements, alpha elements and proton-capture elements
 (those that can be altered in the CNO, NeNa and MgAl nuclear burning cycles),
 respectively.
The numbers in parentheses are the line-to-line scatter for each element
 while the numbers in the super- and subscripts give the estimated ranges
 based on variations in the models.
We include these latter estimates of uncertainties so that our data can
 be compared with other observations that use photometric-based atmospheres
 in the abundance analysis derivations.
Clearly, the uncertainty in the abundance determination is dominated more by
 the uncertainty in the models than by the line-to-line scatter, which we
 use here as the ``error''  under the assumption that our models are
 well-determined.

The Fe-peak elements are consistent with the solar ratio with the exception
 of nickel, which seems to be under-abundant in all the giants by 0.22 ${\pm}$
 0.03 dex on average.
This would be expected if the oscillator strengths for the Ni I lines were
 overestimated by a similar amount, as indicated in Table 3, which
 shows our derived oscillator strengths to be 0.20 ${\pm}$ 0.10 dex larger
 than the literature value.
Thus, using the oscillator strengths from the literature would put [Ni/H]
 closer to zero in our sample.
Why it should be that the oscillator strengths for Ni would be inconsistent
 with the published values while those of the other elements are more
 agreeable remains uncertain.
In fact the difference in {\lgf}'s is even larger since our assumed solar
 Ni abundance is 0.07 dex higher than what \citet{Thev90} assumes; to
 force agreement with the lower Ni abundance value would cause us to increase
 our {\lgf}'s by another 0.07 dex.

\begin{deluxetable}{lcccccc}
\tabletypesize{\footnotesize}
\tablewidth{0pc}
\rotate
\tablenum{7a}
\tablecaption{Fe-Peak Abundances}
\tablehead{
   \colhead{Star}                  &
   \colhead{[Fe/H]$_{\rm Fe I}$}   &
   \colhead{[Fe/H]$_{\rm Fe II}$}  &
   \colhead{[Fe/H]$_{\rm av}$}     &
   \colhead{[V/Fe]}                &
   \colhead{[Mn/Fe]}               &
   \colhead{[Ni/Fe]}
}

\startdata
\cutinhead{M~3}
AA      & $-1.59(0.09)^{+0.11}_{-0.20}$ & $-1.53(0.08)^{+0.08}_{-0.11}$ & $-1.57(0.12)^{+0.12}_{-0.22}$ & $+0.25(0.14)^{+0.17}_{-0.36}$ & $-0.02(0.16)^{+0.17}_{-0.26}$ & $-0.22(0.15)^{+0.15}_{-0.25}$ \\[0.in]
MB 4    & $-1.55(0.07)^{+0.07}_{-0.11}$ & $-1.53(0.06)^{+0.06}_{-0.14}$ & $-1.55(0.09)^{+0.08}_{-0.12}$ & $+0.19(0.11)^{+0.14}_{-0.13}$ & $-0.03(0.14)^{+0.15}_{-0.16}$ & $-0.22(0.14)^{+0.14}_{-0.20}$ \\[0.in]
III-28  & $-1.62(0.07)^{+0.13}_{-0.22}$ & $-1.59(0.07)^{+0.09}_{-0.09}$ & $-1.61(0.10)^{+0.12}_{-0.23}$ & $-0.04(0.13)^{+0.25}_{-0.39}$ & $-0.26(0.11)^{+0.18}_{-0.24}$ & $-0.28(0.12)^{+0.16}_{-0.24}$ \\[0.in]
vZ~297  & $-1.57(0.08)^{+0.14}_{-0.09}$ & $-1.49(0.07)^{+0.08}_{-0.07}$ & $-1.55(0.11)^{+0.14}_{-0.11}$ & $+0.09(0.14)^{+0.26}_{-0.31}$ & $+0.04(0.16)^{+0.22}_{-0.17}$ & $-0.25(0.13)^{+0.15}_{-0.14}$ \\[0.in]
vZ~1127 & $-1.44(0.09)^{+0.09}_{-0.40}$ & $-1.50(0.05)^{+0.05}_{-0.15}$ & $-1.45(0.10)^{+0.10}_{-0.39}$ & $+0.19(0.13)^{+0.13}_{-0.63}$ & $-0.06(0.13)^{+0.13}_{-0.41}$ & $-0.25(0.15)^{+0.15}_{-0.39}$ \\[0.in]
vZ~1000 & $-1.49(0.10)^{+0.10}_{-0.24}$ & $-1.50(0.07)^{+0.07}_{-0.20}$ & $-1.49(0.12)^{+0.10}_{-0.24}$ & $+0.13(0.15)^{+0.15}_{-0.37}$ & $-0.14(0.14)^{+0.14}_{-0.27}$ & $-0.24(0.21)^{+0.21}_{-0.32}$ \\
\cutinhead{M~13}
L~324   & $-1.68(0.08)^{+0.08}_{-0.12}$ & $-1.64(0.07)^{+0.28}_{-0.07}$ & $-1.67(0.11)^{+0.31}_{-0.13}$ & $+0.07(0.14)^{+0.14}_{-0.36}$ & $-0.16(0.11)^{+0.11}_{-0.22}$ & $-0.24(0.15)^{+0.17}_{-0.15}$ \\[0.in]
III-56  & $-1.72(0.07)^{+0.14}_{-0.08}$ & $-1.64(0.03)^{+0.05}_{-0.03}$ & $-1.70(0.08)^{+0.12}_{-0.10}$ & $+0.14(0.11)^{+0.25}_{-0.15}$ & $-0.09(0.08)^{+0.16}_{-0.09}$ & $-0.17(0.12)^{+0.17}_{-0.13}$ \\[0.in]
L~262   & $-1.61(0.07)^{+0.07}_{-0.14}$ & $-1.62(0.08)^{+0.10}_{-0.08}$ & $-1.61(0.11)^{+0.09}_{-0.14}$ & $+0.15(0.14)^{+0.14}_{-0.28}$ & $-0.11(0.14)^{+0.14}_{-0.22}$ & $-0.18(0.14)^{+0.14}_{-0.20}$ \\
\enddata
\end{deluxetable}

Figure 5 shows the Fe-peak abundances as a function of {\teff} for M~3 and M~13.
The [Fe/H] ratios for M~13 are 0.12 dex lower on average than those for M~3;
 although, the difference is only at the 1.5~${\sigma}$ level.
Despite the marginal disparity in [Fe/H], the trends for [V/Fe], [Mn/Fe] and
 [Ni/Fe] are the same for each cluster.

\epsscale{1.00}
\begin{figure}
\plotone{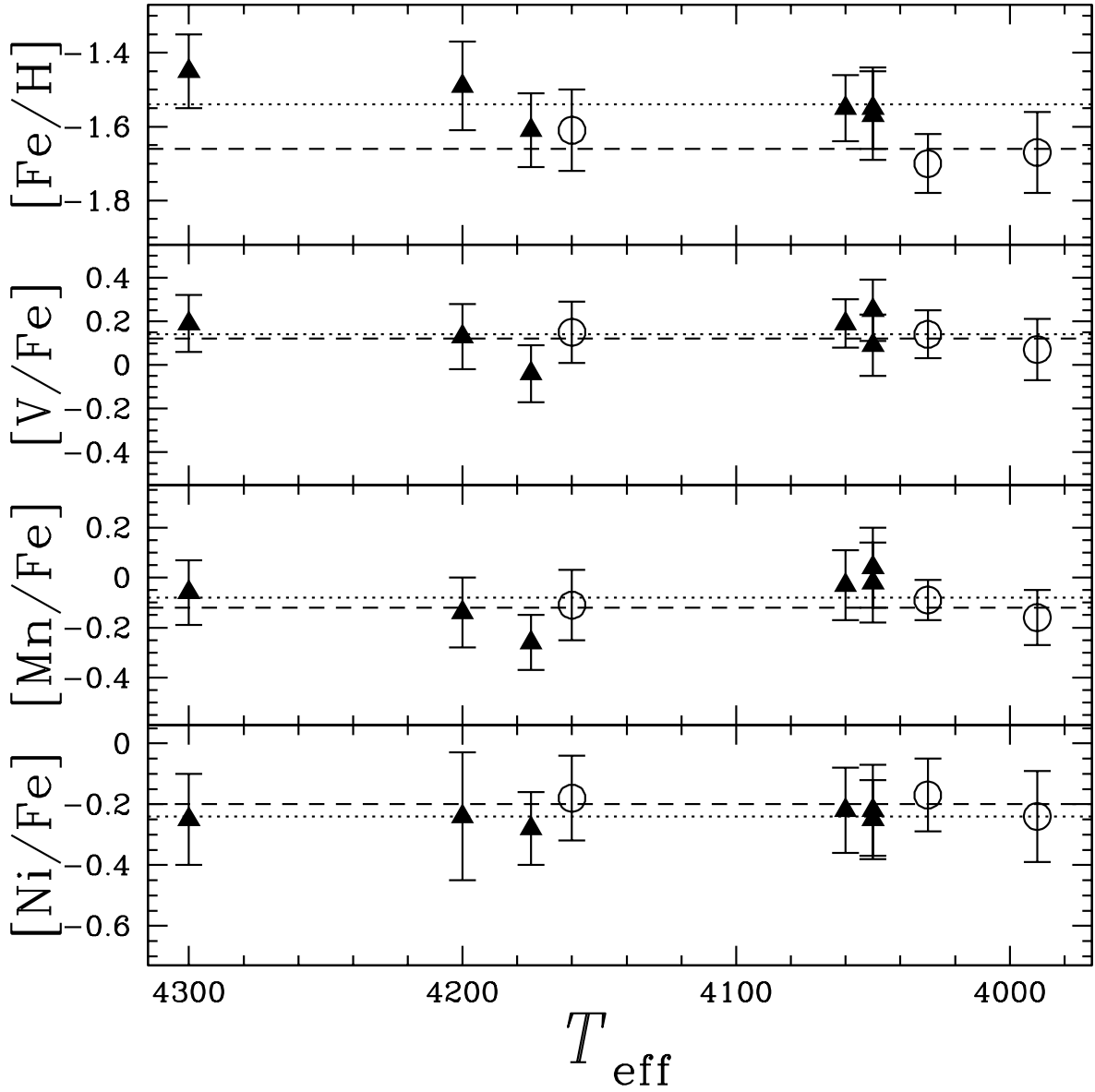}
\caption[Cavallo.fig05.eps]{Fe-peak abundances as a function of {\teff} 
 in M~3 (filled triangles) and M~13 (open circles).
 The dotted and dashed lines are the means for each element for
 M~3 and M~13, respectively, while the error bars are representative of
 the line-to-line scatter. }
\end{figure}

Figure 6 shows the alpha elements as a function of {\teff} for M~3
 and M~13.
The alpha element enhancements are consistent with other observations:
 $[{\alpha}/{\rm Fe}]~=~+0.28 ~{\pm} ~0.08$ for M~3 and $+0.30 ~{\pm} ~0.07$
 for M~13.
The low dispersions probably indicate that the upper and lower limits 
 from the various model atmospheres overestimate the actual errors in the
 abundances.
The two clusters have very similar alpha enhancements, despite the differences
 in their iron abundances.

\begin{deluxetable}{lcccccc}
\tabletypesize{\footnotesize}
\tablewidth{0pc}
\tablenum{7b}
\tablecaption{${\alpha}$ Abundances}
\tablehead{
   \colhead{Star}                  &
   \colhead{[Fe/H]$_{\rm av}$}     &
   \colhead{[Si/Fe]}               &
   \colhead{[Ca/Fe]}               &
   \colhead{[Ti/Fe]$_{\rm Ti I}$}  &
   \colhead{[Ti/Fe]$_{\rm Ti II}$} &
   \colhead{[Ti/Fe]$_{\rm av}$}
}

\startdata
\cutinhead{M~3}
AA      & $-$1.57 & $+0.42(0.14)^{+0.14}_{-0.15}$ & $+0.22(0.13)^{+0.14}_{-0.23}$ & $+0.42(0.15)^{+0.18}_{-0.34}$ & $+0.40(0.12)^{+0.12}_{-0.21}$ & $+0.42(0.19)^{+0.18}_{-0.34}$ \\[0.in]
MB 4    & $-$1.55 & $+0.20(0.12)^{+0.12}_{-0.16}$ & $+0.26(0.12)^{+0.14}_{-0.13}$ & $+0.45(0.11)^{+0.13}_{-0.13}$ & $+0.28(0.10)^{+0.10}_{-0.18}$ & $+0.36(0.15)^{+0.22}_{-0.26}$ \\[0.in]
III-28  & $-$1.61 & $+0.26(0.11)^{+0.12}_{-0.12}$ & $+0.23(0.14)^{+0.21}_{-0.28}$ & $+0.25(0.13)^{+0.25}_{-0.37}$ & $+0.29(0.13)^{+0.14}_{-0.18}$ & $+0.26(0.18)^{+0.38}_{-0.24}$ \\[0.in]
vZ~297  & $-$1.55 & $+0.37(0.16)^{+0.16}_{-0.17}$ & $+0.22(0.14)^{+0.20}_{-0.15}$ & $+0.26(0.15)^{+0.26}_{-0.16}$ & $+0.21(0.11)^{+0.13}_{-0.11}$ & $+0.25(0.19)^{+0.27}_{-0.15}$ \\[0.in]
vZ~1127 & $-$1.45 & $+0.17(0.12)^{+0.12}_{-0.17}$ & $+0.21(0.13)^{+0.13}_{-0.41}$ & $+0.40(0.11)^{+0.11}_{-0.57}$ & $+0.35(0.10)^{+0.10}_{-0.29}$ & $+0.39(0.15)^{+0.12}_{-0.56}$ \\[0.in]
vZ~1000 & $-$1.49 & $+0.29(0.13)^{+0.13}_{-0.15}$ & $+0.28(0.13)^{+0.13}_{-0.26}$ & $+0.29(0.14)^{+0.14}_{-0.35}$ & $+0.20(0.17)^{+0.17}_{-0.23}$ & $+0.27(0.22)^{+0.16}_{-0.33}$ \\
\cutinhead{M~13}
L~324   & $-$1.67 & $+0.31(0.12)^{+0.21}_{-0.12}$ & $+0.21(0.13)^{+0.13}_{-0.28}$ & $+0.34(0.13)^{+0.13}_{-0.32}$ & $+0.24(0.11)^{+0.22}_{-0.11}$ & $+0.32(0.17)^{+0.15}_{-0.30}$ \\[0.in]
III-56  & $-$1.70 & $+0.31(0.11)^{+0.11}_{-0.11}$ & $+0.27(0.12)^{+0.20}_{-0.12}$ & $+0.28(0.11)^{+0.26}_{-0.12}$ & $+0.48(0.09)^{+0.16}_{-0.09}$ & $+0.32(0.14)^{+0.32}_{-0.16}$ \\[0.in]
L~262   & $-$1.61 & $+0.45(0.13)^{+0.14}_{-0.13}$ & $+0.22(0.14)^{+0.14}_{-0.21}$ & $+0.29(0.12)^{+0.12}_{-0.25}$ & $+0.40(0.11)^{+0.11}_{-0.14}$ & $+0.31(0.16)^{+0.20}_{-0.27}$ \\
\enddata
\end{deluxetable}

\begin{figure}
\plotone{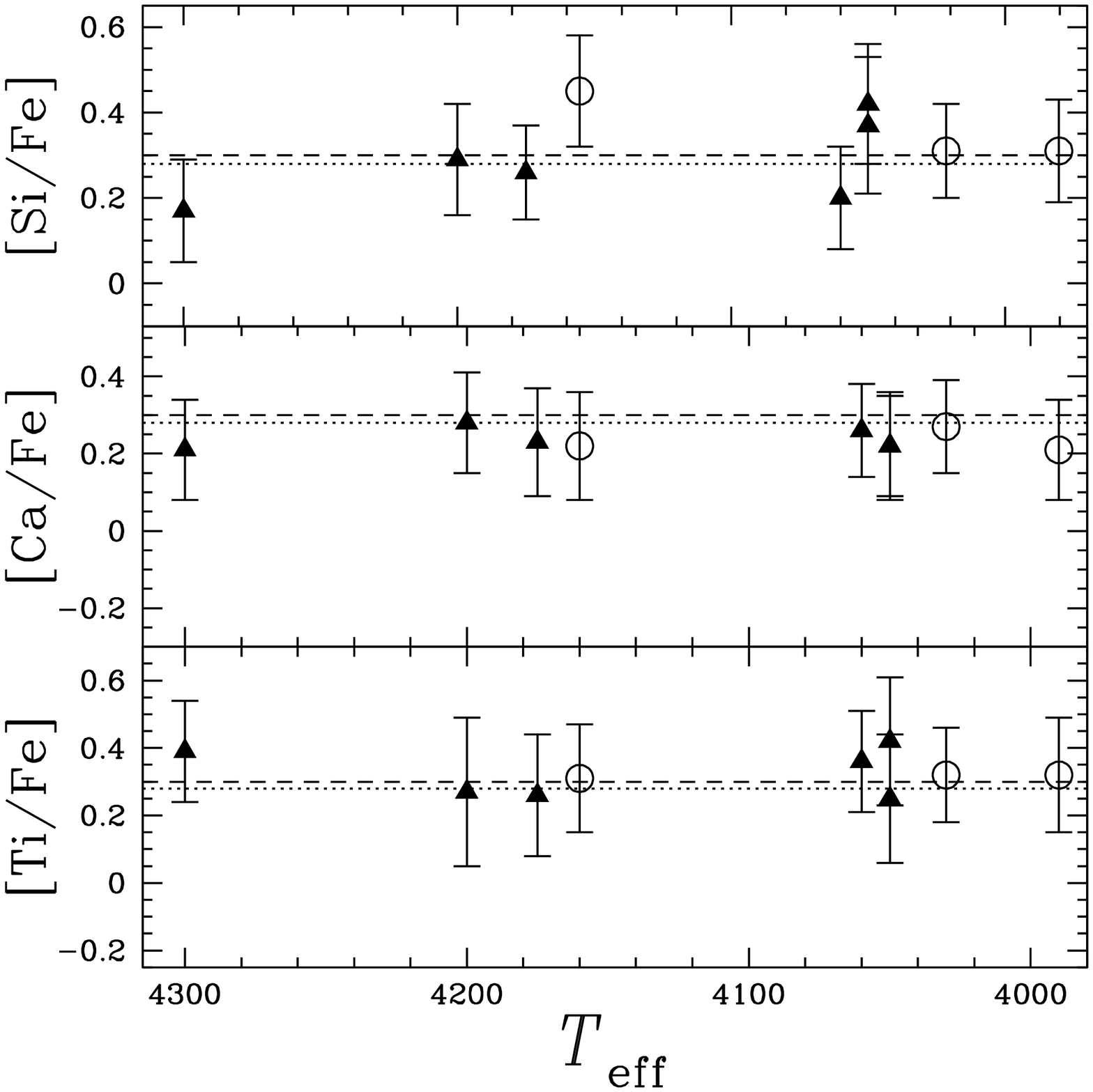}
\caption[Cavallo.fig06.eps]{The alpha elements as a function of {\teff}
 in M~3 (filled triangles) and M~13 (open circles).
 The dotted and dashed lines are the means for each element for
 M~3 and M~13, respectively, while the error bars are representative of
 the line-to-line scatter. }
\end{figure}

Figure 7 shows the trends in the proton-capture element abundances
 for all observed stars in M~3 and M~13 as a function of {\teff}.
The oxygen abundances were calculated via the synthetic-spectra fitting 
 package in MOOG and are presented without error bars because 1) they come
 only from the [O I] line at ${\lambda}$6300 {\AA} so there is no
 line-to-line scatter, and 2) the spectral resolution around the line
 is too low to accurately deblend the line from the nearby Sc II line,
 making attempted variations in the models less meaningful.
We estimate the error in [O/Fe] to be ${\sim}$ 0.15 dex.
The magnesium abundances are derived from only three lines, all near the
 edges of the CCD where the noise is high, and are susceptible to large
 uncertainties.
For the M~3 star III-28, we fit a synthetic spectrum to the data around
 the {\LL}6696,6698 and {\LL}7835,7836 Al~I regions to determine the
 abundance and estimate the error to be ${\pm}$0.15 dex.

\begin{deluxetable}{lccccl}
\tabletypesize{\footnotesize}
\tablewidth{0pc}
\tablenum{7c}
\tablecaption{Proton-Capture Abundances}
\tablehead{
   \colhead{Star}                  &
   \colhead{[Fe/H]$_{\rm av}$}     &
   \colhead{[O/Fe]}                &
   \colhead{[Na/Fe]}               &
   \colhead{[Mg/Fe]}               &
   \colhead{[Al/Fe]}
}

\startdata
\cutinhead{M~3}
AA      & $-$1.57 & $-$0.03 & $+0.42(0.13)^{+0.14}_{-0.19}$ & $+0.25(0.12)^{+0.13}_{-0.14}$ & $+0.87(0.14)^{+0.14}_{-0.20}$ \\[0.in]
MB 4    & $-$1.55 & $-$0.05 & $+0.42(0.10)^{+0.13}_{-0.10}$ & $+0.18(0.13)^{+0.13}_{-0.14}$ & $+0.83(0.13)^{+0.14}_{-0.13}$ \\[0.in]
III-28  & $-$1.61 & $+$0.36 & \nodata & $-0.01(0.11)^{+0.13}_{-0.18}$ & $-0.19(0.15)$ \\[0.in]
vZ~297  & $-$1.55 & $-$0.01 & $+0.13(0.15)^{+0.19}_{-0.16}$ & $+0.05(0.11)^{+0.14}_{-0.11}$ & $+0.71(0.13)^{+0.17}_{-0.13}$ \\[0.in]
vZ~1127 & $-$1.45 & $+$0.33 & $-0.22(0.10)^{+0.10}_{-0.28}$ & $+0.01(0.13)^{+0.13}_{-0.25}$ & $-0.01(0.11)^{+0.11}_{-0.28}$ \\[0.in]
vZ~1000 & $-$1.49 & $-$0.01 & $+0.17(0.12)^{+0.13}_{-0.20}$ & $+0.18(0.13)^{+0.13}_{-0.19}$ & $+0.72(0.14)^{+0.14}_{-0.21}$ \\
\cutinhead{M~13}
L~324   & $-$1.67 & $-$0.38 & $+0.54(0.12)^{+0.12}_{-0.23}$ & $-0.02(0.13)^{+0.13}_{-0.15}$ & $+0.99(0.12)^{+0.12}_{-0.19}$ \\[0.in]
III-56  & $-$1.70 & $-$0.05 & $+0.50(0.08)^{+0.13}_{-0.09}$ & $+0.23(0.16)^{+0.19}_{-0.17}$ & $+0.74(0.11)^{+0.16}_{-0.11}$ \\[0.in]
L~262   & $-$1.61 & $+$0.11 & $+0.34(0.11)^{+0.11}_{-0.16}$ & $+0.10(0.14)^{+0.14}_{-0.17}$ & $+0.61(0.13)^{+0.13}_{-0.17}$ \\
\enddata
\end{deluxetable}

In M~3 [Al/Fe] spans a range of 1.1 dex over 250 K and, while the three
 cooler stars appear to be more enhanced than the three hotter ones on
 average, it must be cautioned that the sample is biased since the stars
 were chosen from the Lick/Texas studies based on evidence for or against
 mixing and is not close to being complete.
In addition, the oxygen abundances exhibit a strong anticorrelation with
 aluminum, but aren't as depleted as in the so-called 
 ``super-oxygen-poor'' giants in M~13 \citep{KSLP92}.
The sodium abundances likewise show an increasing trend with decreasing
 {\teff} and are correlated with [Al/Fe].
Finally, [Mg/Fe] seems fairly independent of {\teff} and doesn't appear
 to show any correlations with the other proton-capture elements.

In M~13 it is impossible to make any firm conclusions since the
 data are so few; however, we note several trends.
First, [O/Fe] and [Al/Fe] are strongly anticorrelated.
Second, the aluminum abundance is high ($>$ 0.6 dex) for all three stars,
 while the oxygen varies from slightly enhanced to fairly depleted 
(but again, not super oxygen poor).
Third, [Na/Fe] shows a slight trend of increasing with decreasing {\teff}.
Fourth, the one giant with the strongest aluminum enhancement (L~324) has
 the strongest magnesium depletion and is also the brightest star in the sample.

\begin{figure}
\plotone{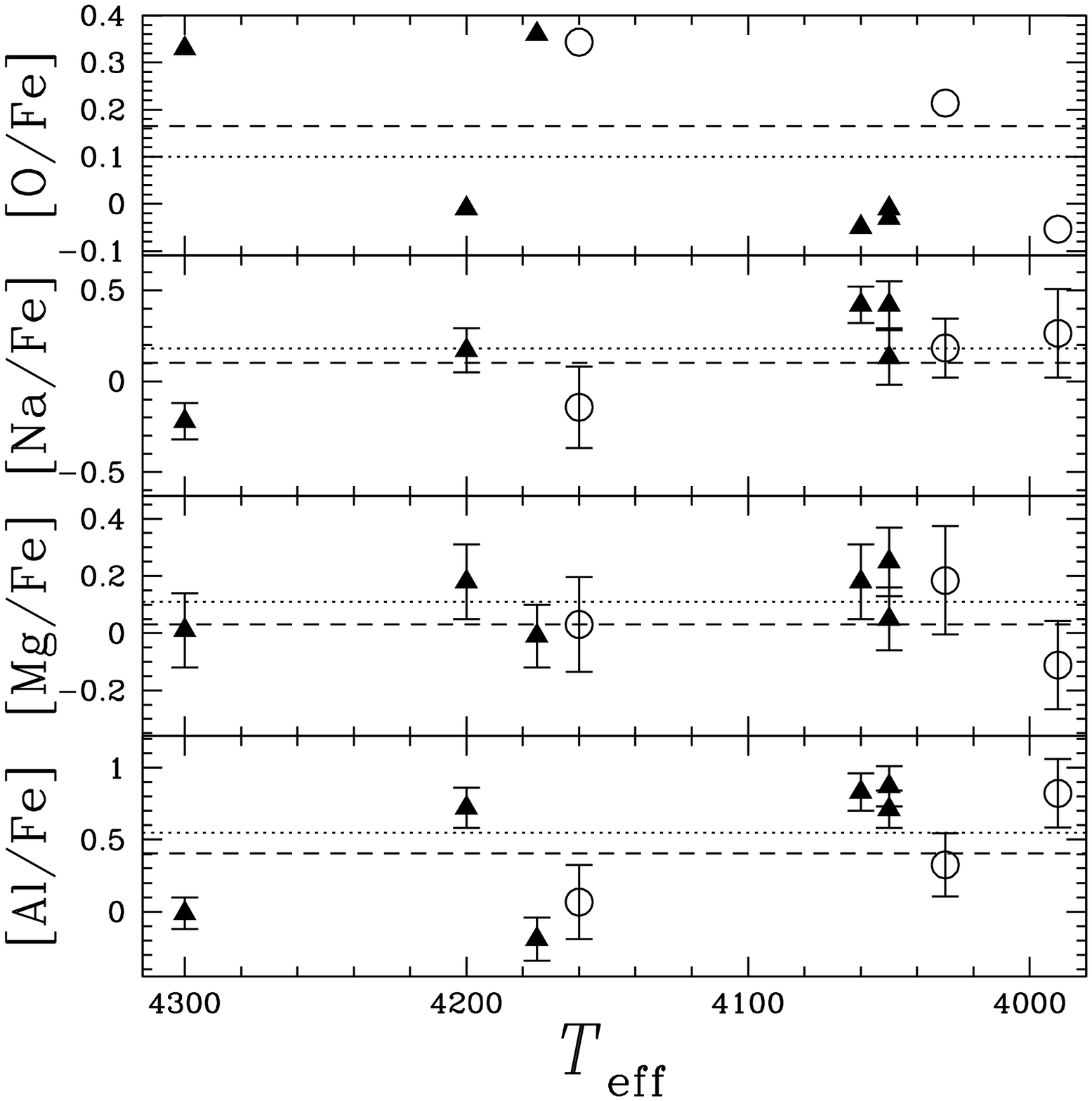}
\caption[Cavallo.fig07.eps]{The proton-capture elements as a function of
 {\teff} in M~3 (filled triangles) and M~13 (open circles). 
 The dotted and dashed lines are the means for each element for
 M~3 and M~13, respectively, while the error bars are representative of
 the line-to-line scatter. }
\end{figure}

\section{DISCUSSION}
\subsection{Evidence for Deep Mixing on the RGB}
\subsubsection{Theoretical predictions}

In \citet{CSB96} and CSB98 we explored the development of the abundance
 profiles around the H~shell of four canonical stellar evolutionary sequences.
Although the models were unmixed, we can infer some predictions with regard to
 deep mixing, which, we remind the reader, is defined as mixing that
 penetrates the H shell.
\begin{itemize}
\item Carbon, nitrogen and oxygen are not good tracers of deep mixing since 
 they are easily altered above the H~shell in the CN and ON nuclear reaction 
 cycles.
\item Sodium is altered above the H~shell from $^{22}$Ne and inside the
 H~shell from $^{20}$Ne through the NeNa cycle as shown in Figure 8.
 The proton-capture rates for the NeNa cycle are still uncertain and the
 initial neon abundance in real RGB stars is impossible to measure,
 making the theoretical prediction of actual sodium enhancements difficult. 
\item As shown in Figure 8, $^{27}$Al, the only stable aluminum isotope, 
 is enhanced only inside the H~shell at the expense of mostly $^{25}$Mg plus
 $^{26}$Mg, but also some $^{24}$Mg deep inside the H~shell
 for very bright, metal-poor models.  The reaction rates for the MgAl cycle
 are still subject to large uncertainties; although, the $^{24}$Mg proton
 capture rates are now well-determined \citep{Powell99}.
\item Aluminum enhancements are temperature sensitive, indicating that they 
 should not be expected until higher luminosities are achieved in lower 
 metallicity giants ([Fe/H]~${\lesssim}~-1.2$).
 In addition, the creation of sodium from $^{20}$Ne also requires the high
 temperatures found only inside the H~shell of the same bright, metal-poor
 giants, indicating that large sodium enhancements that originate with
 this neon isotope occur only towards the RGB tip.
\end{itemize}

\begin{figure}
\plotone{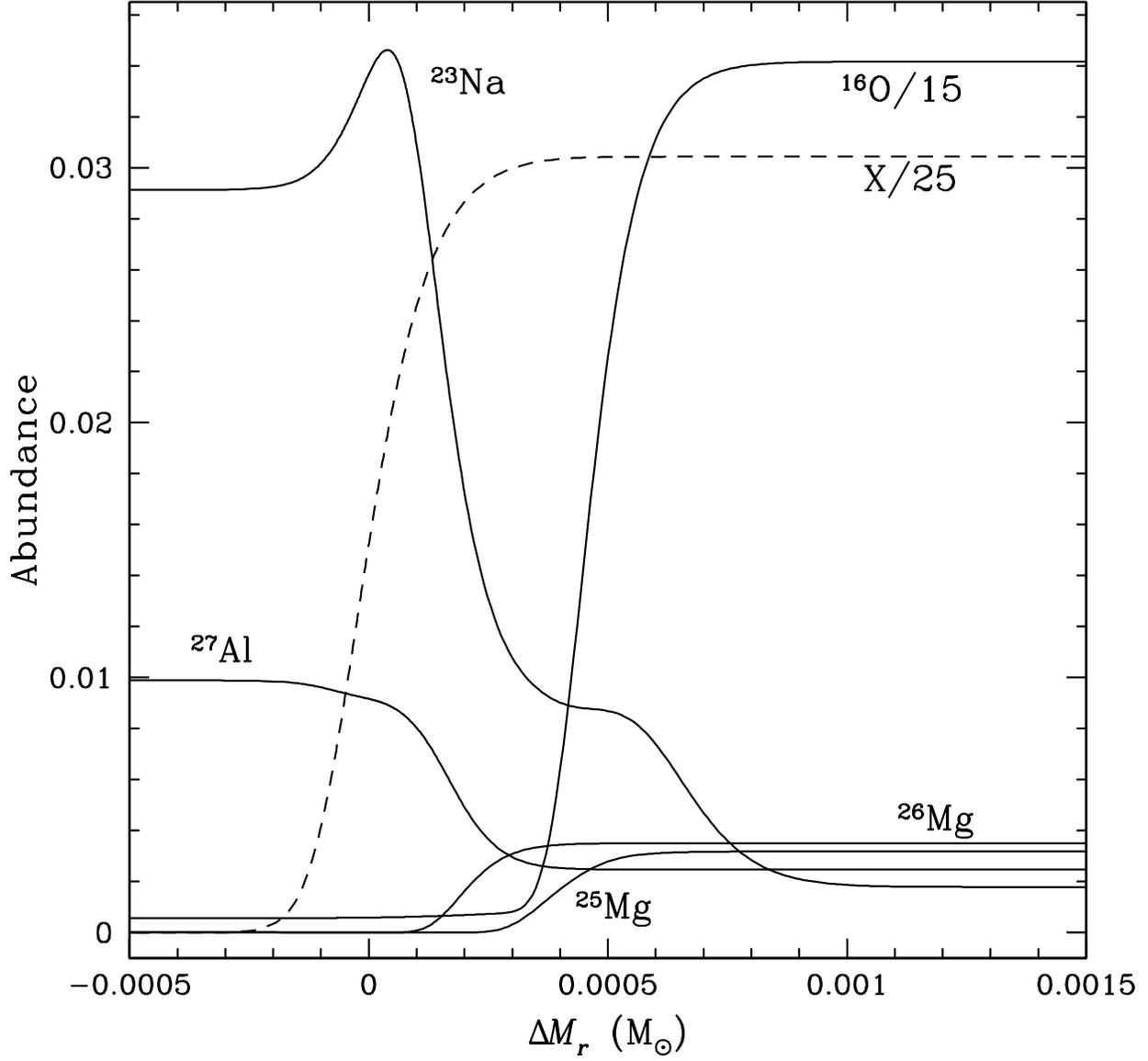}
\caption[Cavallo.fig08.eps]{The abundance variations around the H shell of a
 red-giant-branch tip model with [Fe/H]~=~$-1.67$ and scaled solar
 composition.
 The abscissa is the mass difference between any point and the
  center of the H shell.
 Hydrogen and helium are given in mass fractions while the other elements
  are scaled relative to the total number of metals.}
\end{figure}

All these inferences are still subject to uncertainties in
 the source of mixing, the initial abundances and the nuclear 
 reaction rates; nonetheless, we now venture to compare the observational 
 data with the theoretical predictions.

\subsubsection{Observational results: the aluminum data}

In addition to the M~13 aluminum abundances determined above, we were
 provided with the equivalent-width data of the Al~I ${\lambda}$6696 {\AA}
 line for 78 more giants in this cluster that were obtained from spectra
 taken with the WIYN telescope and Hydra multi-object spectrograph by 
 Dr. C. Pilachowski in an attempt to find spurious Li~I
 features at ${\lambda}$6707 {\AA} \citep{PSKHW2000}, as found in the M~3
 giant IV-101 by \citet{Kraft99}\footnote{We find no evidence of this feature
 in any of our M~3 or M~13 spectra.}.
The details of the data reduction and search results for her study
 will be reported on elsewhere \citep{PSKHW2000}.
Since the Li~I line is so strong, the exposures were short in order
 to probe as many stars as possible.
Unfortunately, this resulted in a lower signal-to-noise ratio than preferred
 for the nearby Al~I lines, but the data are reliable for stars with
 strong aluminum lines: of the 78 giants in the WIYN dataset, 66 had
 measurable equivalent widths.
To derive the aluminum abundances, the equivalent widths were folded through
 models that were built using the MARCS code based on the atmospheric parameters
 that were initially derived via photometry \citep{PSKL96}.
The models were assumed to have [Fe/H]~=~$-1.50$, while, as with the models
 from this present study, the [${\alpha}$/Fe] ratio was also assumed to be 
 enhanced by $+$0.4 dex.
As shown in Tables 5, 6 and 7, models based on photometric indices can
 lead to a wider range of abundances; although, we believe the
 WIYN models to be well-determined since they were iteratively corrected
 with the spectra.

Of the twelve other stars in the WIYN sample that did not have measurable
 lines, one, II-76, had an aluminum abundance previously determined in the
 literature; [Al/Fe]~=~$-0.19$ \citep{S96a}.
The other eleven were assumed to have [Al/Fe]~=~0 for the purposes of the
 statistics discussed below, an assumption that seems verified by II-76
 having such a low aluminum abundance; although, this star is rather
 bright while the other eleven are much lower magnitude examples.
The three M~13 stars observed by us were also present in the WIYN sample and
 the abundances agreed to within the errors from the line-to-line scatter.
Although the differences between the 4-m data and the WIYN data are small,
 systematic errors can arise since the WIYN models are rooted in photometry
 with corrections to {\teff} and $v_t$ from lower resolution spectra
 (see Pilachowski et al. 1996 for details).
For example, our models are 40-60~K hotter, have gravities that are
 0.35-0.45 dex lower and microturbulent velocities that are 0.09-0.13
 km~s$^{-1}$ higher.
In addition, our spectra have a factor of two higher resolution than the
 WIYN data and our signal-to-noise ratios are significantly higher.

Finally, the sample was then augmented with [Al/Fe] values taken from the
 literature \citep{WLO87,S96a,Kraft97}, bringing the total number of stars
 with determined [Al/Fe] values to 85.
We believe that the systematic errors that might be present in the
 data are mostly removed before combination since, with the exception
 of \citet{WLO87} for two stars, they are derived by the same group of
 Lick/Texas observers who practice consistent reduction techniques.
Small differences will arise as the telescopes and instruments are varied,
 but the Lick/Texas observers do compare their various observations
 and show little scatter among their results.
If one adds the eleven stars with the assumed low [Al/Fe] values, the total
 sample size is 96, covering a range from the tip of the RGB at $V~=~11.9$ to
 $V~=~15.5$, with complete coverage of the \citet{CM79} photometry for
 $V~{\lesssim}~13.7$.
The data are plotted in Figure 9a as a function of $V$ magnitude, where
 the eleven stars with [Al/Fe] assumed to be 0 are shown as open circles.
The [Al/Fe] values of giants with multiple measurements were averaged
 together after being normalized by [Fe/H].

\begin{figure}
\plotone{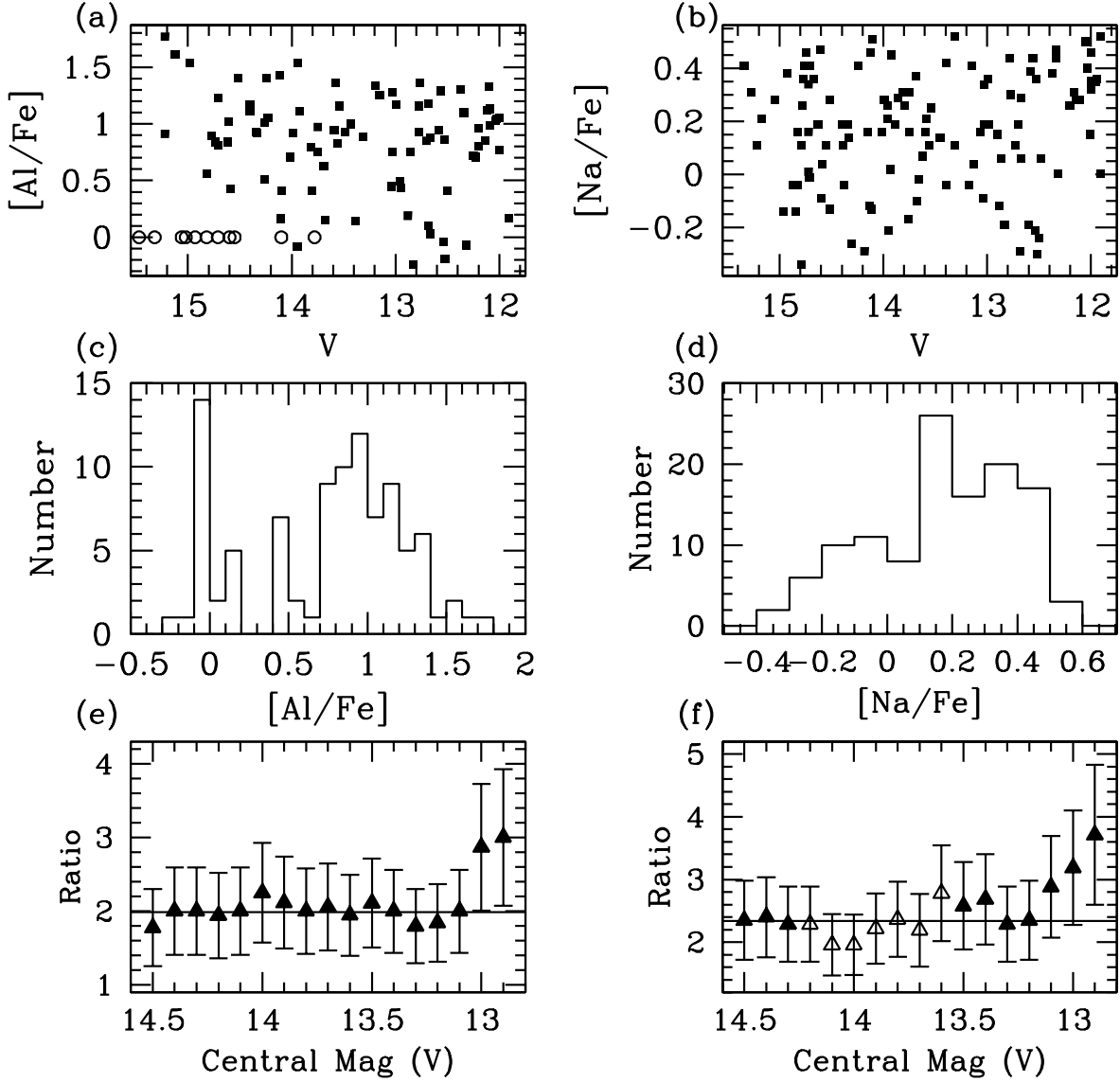}
\caption[Cavallo.fig09.eps]{
(a) [Al/Fe] values of M~13 giants as a function of $V$ magnitude.
 Measured values are shown as filled squares, while those with assumed
 values are shown as open circles.  In the cases where multiple measurements
 for the same star exist, the values are averaged together after being
 normalized by [Fe/H]. 
(b) Same as (a), except for [Na/Fe].
(c) A histogram of the [Al/Fe] distribution in M~13.
(d) Same a (c) except for [Na/Fe].
(e) The Al ratio in M~13 as a function of magnitude as determined by
    the KMM algorithm.
    The thick solid line is the mean of the 15 points at $V~>~13.0$.
(f) The Na ratio in M~13 as a function of magnitude as determined by
    the KMM algorithm. 
    The open triangles indicate where the distribution is likely unimodal.
    The thick solid line is the mean of the 14 points at $V~>~13.1$.}
\end{figure}

In Figure 9c we show a histogram of the distribution of [Al/Fe] for the 
 entire sample of 96 M~13 giants.
From this figure, we see an apparent gap between [Al/Fe]~=~0.2 and 0.4,
 indicating that the distribution might be bimodal.
We test for bimodality by applying the KMM algorithm of \citet{KMM}, which
 tests the null hypothesis that a single Gaussian is a good description of
 the data by comparing the fit of a single-peaked distribution to one
 with multiple modes.
The algorithm returns a $P$-value that describes the confidence level of 
 the single-mode fit, where  $P~<~0.05$ indicates that a single Gaussian
 can be rejected at better than the 95\% confidence level, which is generally
 accepted as strongly consistent with the multimodal distribution.
Testing for a bimodal distribution in our sample gives $P~=~0.000$ with
 means in [Al/Fe] of $0.12~{\pm}~0.25$ for the ``Al-normal'' peak and
 $1.03~{\pm}~0.25$ for the ``Al-enhanced'' peak.
The number of stars in each distribution is 65 and 31 for the Al-enhanced
 and Al-normal groups, respectively, giving a ratio of Al-enhanced
 to Al-normal stars (hereafter, the ``Al ratio'') of $2.10~{\pm}~0.46$,
 where the error is estimated from Poissonian statistics.

It is possible that our assumption concerning the actual [Al/Fe] values
 of stars with no measurable Al~I lines can introduce a bias into our
 statistics.
For example, \citet{PSKHW2000} report an upper limit of 20~m{\AA} for 
 equivalent width measurements of lower luminosity giants.
Applying this measurement to the star K~272, which has $V~=~15.47$, and
 using the model parameters supplied by \citet{PSKHW2000}, we obtain 
 [Al/Fe]~${\leq}$~0.88, assuming [Fe/H]~=~$-$1.49.
We test the effect of this bias by subjecting just the 85 giants with actual
 aluminum measurements to the KMM algorithm,
 yielding a $P$-value of 0.002, with an Al-ratio of $3.72~{\pm}~0.98$.
This clearly demonstrates that removal of the uncertain data still results in 
 a strongly bimodal distribution.
The only real solution to correct this possible bias is to make higher
 signal-to-noise observations.

If deep mixing is occurring on the RGB, then the Al ratio should be a
 function of magnitude, increasing with decreasing $V$.
To test this hypothesis we bin the data by magnitude and apply the KMM algorithm
 to each bin to determine whether the distribution in each bin is bimodal and,
 if it is, the Al ratio.
The KMM algorithm requires that the number of data points be greater than
 50, forcing the size of magnitude bins to be rather wide (${\Delta}V~=2$)
 in order to ensure that enough stars are included for reliable statistics.
We began our binning at $V~=~12.9$, one magnitude lower than the brightest
 star in the sample, 
 and shift each bin by 0.1 magnitude up to $V~=~14.5$, one magnitude
 brighter than the lowest luminosity star in the sample.
This choice of bins avoids adding empty points along the RGB into our 
 statistics; although, it reaches magnitudes where the sample is incomplete.
According to the KMM algorithm, the aluminum distribution in each magnitude
 bin is bimodal, with all $P$ values less than 0.013 and most less than 0.004.
In Figure 9e we show the Al ratio (with Poissonian error bars)
 as a function of the central magnitude of each bin.
The mean of the 15 points between $V~=~13.1$ and 14.5 is 1.99,
 which is shown as the solid horizontal line in Figure 9e.
The Al ratios in the second brightest and brightest magnitude bins are
 2.9 and 3.
The upturn at the brighter magnitudes is due to both an increase in the number
 of Al-enhanced stars and a decrease in the number of Al-normal stars, as
 can be seen in Figure 9a, and is consistent with mixing occurring along
 the RGB; although, the error bars do not allow for a definitive conclusion
 in this regard.

We now compare the M~13 data with those from M~3, where the number of
 giants with measured Al abundances is substantially smaller.
Although we augment our sample of six giants in M~3 with an additional
 four that were observed by \citet[none in common with our sample]{Kraft99},
 the numbers are still too small to apply the KMM algorithm; however,  
 as shown in Figure 10a, the M~3 [Al/Fe] distribution
 appears bimodal, with an Al ratio of $1.5~{\pm}~1.0$.
This is consistent with the presumably unmixed dimmer giants in M~13, 
 suggesting that deep mixing is not occurring in M~3.

\epsscale{0.50}
\begin{figure}
\plotone{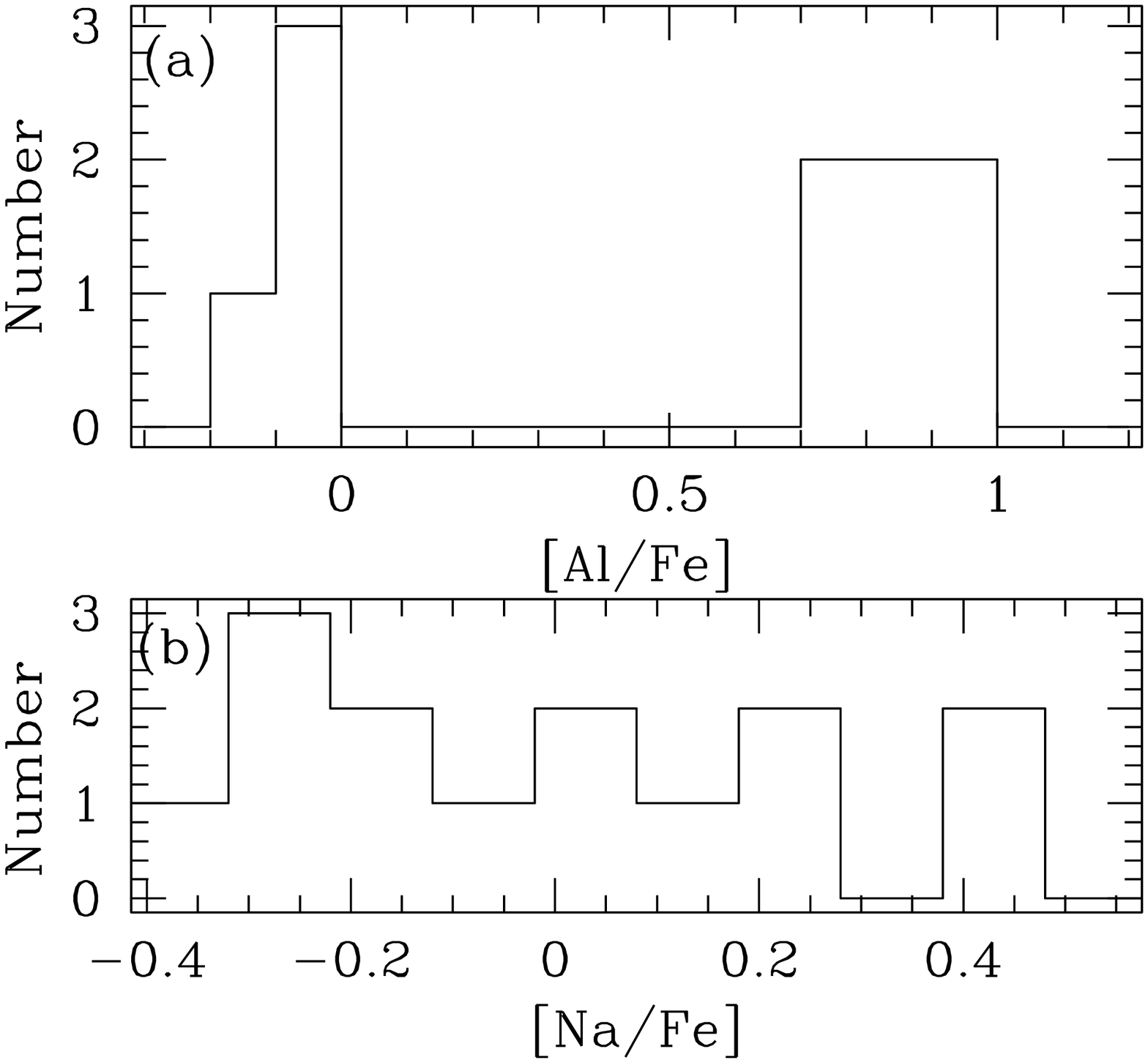}
\caption[Cavallo.fig10.eps]{A histogram of (a) the [Al/Fe] and (b) the 
 [Na/Fe] distributions in M~3.}
\end{figure}

\subsubsection{Observational results: the sodium data}

We incorporated the [Na/Fe] data from our three M~13 giants with the literature
 values \citep{LBC91,KSLP92,KSLS93,KSLSB95,PSKL96,S96a,S96b,Kraft97} to
 build a database containing 119 M~13 RGB stars with measured sodium abundances,
 which we show in Figure 9b as a function of $V$.
Again, we believe the systematic errors associated with the combination of
 various data sets to be minimized for the reasons outlines in the previous
 section.
As demonstrated in Figures 9a and 9b, and in Figure 11 for a subset of stars
 that have both [Al/Fe] and [Na/Fe] values determined, the range in [Na/Fe]
 is not as wide as in the [Al/Fe] data, with the [Na/Fe] values being both
 more negative for ``low'' sodium stars and not as enhanced for the ``high''
 sodium stars.
However, from Figure 9b it is clear that the tip of the RGB does lack
 sodium-poor giants, if [Na/Fe]~=~0 can be considered a high value relative
 to the rest of the low-Na distribution.

A histogram of the total sodium distribution is shown in Figure 9d,
 where two peaks are apparent, but with no obvious gap between them.
However, application of the KMM algorithm does indicate a bimodal
 distribution ($P~=~0.001$) with two peaks at [Na/Fe]~=~$-$0.09~${\pm}$~0.13 
 and $+$0.29~${\pm}~$0.13.
The low sodium group has 33 members, while the high sodium group has 86,
 for a ratio of Na-enhanced to Na-poor giants (hereafter, the ``Na ratio"")
 of 2.61~${\pm}~0.53$.
Since the peaks are not widely separated (2.93 ${\sigma}$), it is
 difficult to determine to which group stars between the peaks belong,
 making the above ratio less certain.
The KMM algorithm provides two group membership probabilities (GMPs) for each 
 star, which give the percentage probability that a value belongs to ``high''
 and ``low'' mode (the sum of the GMPs for each star equals 100\%).
For the sample of M~13 giants, 35 stars (30\%) have both GMPs between 10\%
 and 90\%, indicating that these stars cannot be assigned to either mode with
 high confidence.
 
When we bin the [Na/Fe] values by magnitude, as with the [Al/Fe] data, the
 bimodality of the distribution within each bin is not as certain as with
 the entire sample.
We demonstrate this in Figure 9f, where we show the Na ratio for M~13 in 
 two-magnitude wide bins.
The open triangles in this figure represent ratios where the $P$-values are
 greater than 0.05, indicating that a unimodal Gaussian is not easily
 ruled out as the ``true'' parent distribution.
The filled-in triangles at $V~>~14.2$ represent apparently bimodal 
 distributions; however, the stars at these lower magnitudes are
 undersampled, confounding efforts to determine the nature of the [Na/Fe]
 distribution in M~13.

Despite the uncertainty at lower magnitudes, the [Na/Fe] distribution for the 
 brighter M~13 giants is likely bimodal according the KMM statistics
 and is similar to that of the [Al/Fe] distribution.
The mean of the 14 dimmest points is 2.33, represented by the solid line 
 in Figure 9f.
Relative to the mean, the upturn at $V~=~13.1$ appears real and is due to
 the lack of low-sodium stars $V~{\lesssim}~12.5$, as seen in Figure 9b.

We compare the distribution of sodium in M~3 with that of M~13.
As with the aluminum data, the sample size is small, with only fourteen M~3
 stars having determined [Na/Fe] values \citep{KSLP92,KSLS93,KSLSB95,Kraft99},
 six of which are also determined above.
We believe the systematic differences among the Lick/Texas results and
 between our data and theirs that arise from the use of different telescopes,
 instruments and reduction techniques are not significant enough to affect the
 interpretation of the results.
As can be seen in Figure 10b the distribution is fairly flat with eight stars
 having [Na/Fe]~${\le}~0.00$ and only three with [Na/Fe]~$>~+0.3$.
This is more consistent with limited mixing, but the numbers are much
 too small to draw realistic conclusions.

\subsubsection{Observational results: sodium and aluminum}

Referring again to Figure 11, we look for a correlation between sodium and
 aluminum in M~13 by comparing the 62 giants with measured values of both
 [Al/Fe] and [Na/Fe].
The data appear correlated, with a linear correlation coefficient of 0.723,
 which according the probability coefficient given in Appendix C
 of \citet{Taylor}, is ``highly significant'' for this sample size.
We do note, however, that at [Al/Fe]$~{\sim}~+0.5$, the full range of
 [Na/Fe] values is present, making the correlation suspicious around
 this narrow range of [Al/Fe] values.

\begin{figure}
\plotone{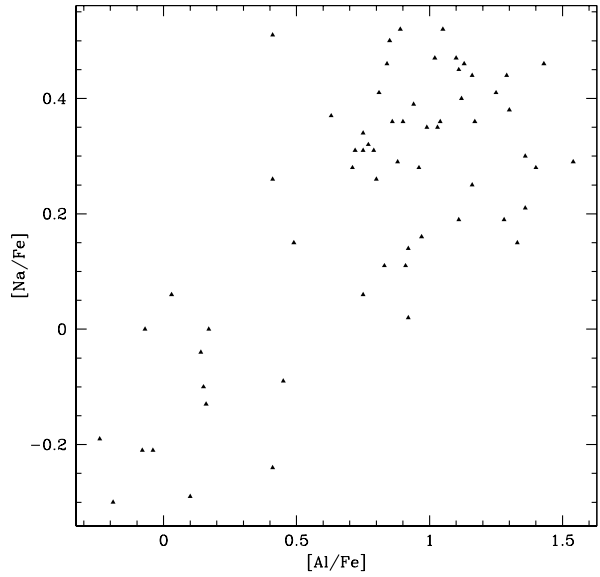}
\caption[Cavallo.fig11.eps]{[Na/Fe] versus [Al/Fe] for 62 giants in M~13.}
\end{figure}

Applying the KMM algorithm to just these 62 stars reveals that both the
 aluminum and sodium distributions are bimodal, with
 each having $P~=~0.000$, and an Al ratio and a Na ratio both equal 
 to 3.13~${\pm}~0.93$.
To test whether the identical ratios are just coincidence or if indeed,
 a star with high [Al/Fe] is likely to have high [Na/Fe] and vice-versa,
 we examine the difference in the GMPs between the aluminum and sodium data
 for the ``high'' modes, as shown in the histogram in Figure 12.

If the abundances of aluminum and sodium are correlated then
 the difference between the GMPs will be close to zero, as seems to be the
 case for most giants since 52 stars fall between $-$0.10 and $+$0.10.
However, ten deviate from zero by more than 0.25; so that while Figure 12 
 indicates that only two or three giants do not fit into the correlation
 around [Al/Fe]~=~$+0.5$, the KMM algorithm actually shows that this number 
 is larger and that around 16\% of the sample are not statistically correlated.
We must also reiterate that the [Al/Fe] values from the WIYN
 sample are not always well-determined and these numbers are likely
 to change with better data.
In general, the correlation between [Al/Fe] and [Na/Fe] seems fairly
 constrained; however, we suggest that when testing for deep mixing, 
 aluminum is a better element to observe than sodium since the data show
 that the distinction between high and low [Al/Fe] is clearer and the
 models suggest that aluminum is produced much closer to the H shell
 than sodium.
According to the models, the appearance of {\em mixing-enhanced} aluminum
 on the surface of a star should imply the existence of extra sodium, 
 while the converse is not necessarily true.
The fact that the data show that 84\% of the time the abundance of one
 element is a predictor of the other indicates just how
 deeply mixed the M~13 giants probably are.

\begin{figure}
\plotone{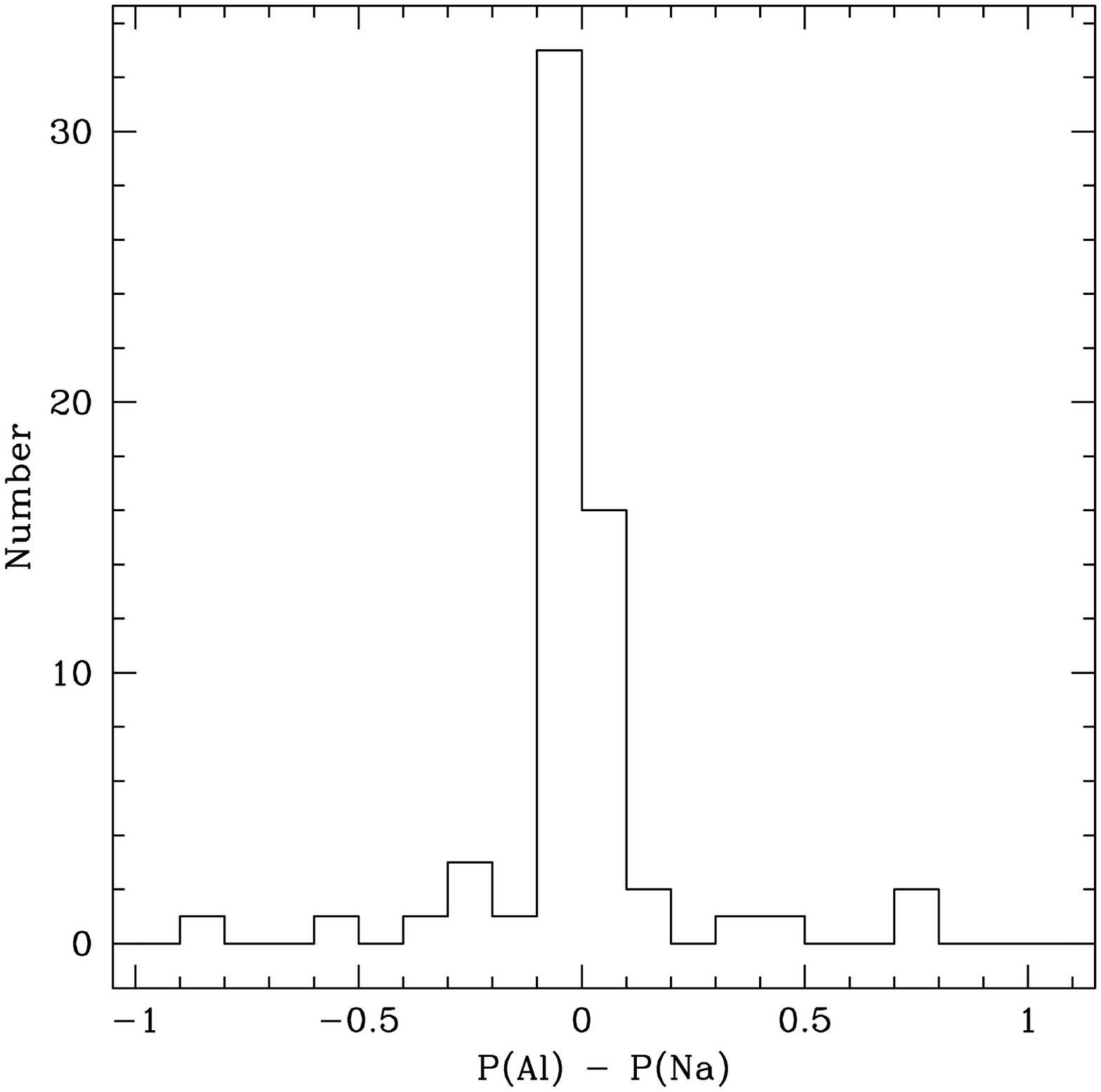}
\caption[Cavallo.fig12.eps]{A histogram of the differences between the group
 membership probabilities for aluminum and sodium for the stars in Figure 11.}
\end{figure}

\subsection{Hot Flashes and Primordial Influences}

Enhanced aluminum abundances at magnitudes much lower than the tip of RGB
 are difficult to explain with mixing models since the peak temperature of
 the H~shell is not high enough to produce a significant amount of
 aluminum at this stage of evolution \citep{CSB98}.
What then is the source of the high aluminum abundances at these lower
 luminosities?
Some have suggested that the H~shell might become unstable at lower 
 magnitudes due to rotation and can undergo flashes that result in peak
 temperatures near 70 MK or higher \citep{LHZ97,FAK99}, as opposed to
 the canonical temperatures below 60 MK (see, e.g., CSB98).
This hot temperature was chosen specifically by \citet{LHZ97} because
 it reproduces the observed abundance anomalies in some M~13 giants,
 particularly the $^{24}$Mg depletions and aluminum enhancements observed
 by \citet{S96b}.
Unfortunately, such an exercise depends strongly on the
 accuracy of the nuclear reaction rates, which, in many cases, are
 not very well determined 
 [see, e.g., the NACRE \citep{NACRE} compilation\footnote{Also available
 at \url{http://pntpm.ulb.ac.be/nacre.htm}}].
Also, in addition to the fact that the H-shell instabilities have yet
 to be demonstrated in RGB models, it is not clear what the effects of such
 flashes would be on the structure and evolution of a star.
For example, would these flashes have observable consequences that affect the
 RGB luminosity function, which is generally well-reproduced by canonical
 evolutionary models? 
Furthermore, the instability scenario of \citet{LHZ97} only applies to the
 lower RGB \citep{VVH88}, which is at odds with the data reported here that
 show aluminum and sodium enhancements occur towards the tip of the RGB;
 if these elements were being produced on the lower RGB via hot flashes,
 the Al and Na ratios would vary at lower magnitudes.
In the case of the \citet{FAK99} scenario, which involves continually peeling
 off layers of the core and completely disrupting the H~shell, it is not clear
 how stars can evolve up the RGB and not have serious consequences for the
 observed luminosity functions of clusters in which the RGB members experience
 deep mixing.

We suggest that a more favorable location for hot hydrogen burning is around
 the H~shell of intermediate-mass ($M~>~4 {\rm M}_{\odot}$) AGB stars 
 (referred to as IMS), which undergo hot bottom burning (HBB), so called because
 the convective envelope is in contact with the H~shell.
The IMS could have shed their nuclearly processed envelopes that include
 enhanced aluminum and sodium abundances into the early cluster environment
 \citep{CD81,DWW97,DDNW98}, creating the observed abundance distributions.
For example, the bimodality of [Al/Fe] values on the lower RGB
 would be created if the ejecta were distributed locally.
Likewise, the [Na/Fe] distribution in both clusters is explained
 if the IMS envelopes were also enriched in sodium.
Unfortunately, detailed and accurate aluminum and sodium abundance yields
 from metal-poor AGB evolution models are non-existent; but
 the high temperatures of HBB in IMS, and the observations themselves,
 lend some weight to this hypothesis.

\subsection{Results from a Deep Mixing Algorithm}

We have taken the models described in CSB98 and subjected them to a deep 
 mixing algorithm that assumes that the mixing is instantaneous; 
 i.e., the mixing timescale is the same as the nuclear burning timescale.
A complete derivation of our algorithm is given in the appendix.
While this simplified approach is unable to mimic a realistic
 mixing timescale, it does have several advantages:
 1) it can give an upper limit to the amount of variation an element can 
    experience due to nuclear processing, 
 2) it can show the lowest point on the RGB where an element can 
    be processed and 
 3) it can be used to check the effect of the uncertainties in the 
    nuclear reaction rates on the envelope abundances \citep{Cavallo98}.
We discuss the first two points after a brief description of the algorithm.

The nuclear reaction network employed in the mixing algorithm is the same as
 the one used in CSB98 with the following modifications.
We use updated rates for the $^{26}$Mg$(p,{\gamma})^{27}$Al reaction
 that have been provided to us by C. Rowland.
Her rates are 1) approximately 10 - 16 times faster than the NACRE rates
 \citep{NACRE}, 2) 1.5 to 4.5 times faster than the rates used in
 CSB98 and 3) commensurate with the \citet{CF88} tables in the range of
 $T_9~=~0.4-0.6$, where $T_9~=~10^{9}$~K.
The $^{26}$Al proton-capture rate has been separated into the short-lived
 isomeric state, $^{26m}$Al, and the meta-stable ground state,
 $^{26g}$Al (this had a negligible impact on the conclusions drawn in CSB98).
The NACRE compilation shows that the $^{26g}$Al proton-capture rate is
 uncertain by three orders of magnitude, the effects of which are 
 discussed in the conclusions. 
We use the rates for the $^{24}$Mg$(p,{\gamma})^{25}$Al reaction that
 have been updated by \citet{Powell99}, who measured the resonance 
 parameters of the E$_{R}$~=~223~keV resonance to show that the low-energy
 contribution to the total rate does not significantly 
 increase this rate as suggested by \citet{ZL97}.
The new rates show a 32\% increase over the commonly used \citet{CF88} rates in
 the range of $T_9~=~0.4-0.6$, which is not enough to account for the observed
 depletions of $^{24}$Mg in a handful of M~13 and NGC~6752 giants observed by
 \citet{S96b,S97}.
 
The initial abundances that we put into our algorithm are those of
 \citet{DDNW98}, who suggest using [$^{25}$Mg/Fe]$~=~+1.1$~dex as the result
 of AGB contamination, while the initial [$^{24}$Mg/Fe] and [$^{26}$Mg/Fe]
 both equal 0.
This suggestion is further backed by recent results of \citet{LFC99} who
 find the overproduction of $^{25}$Mg and $^{26}$Mg relative to
 $^{24}$Mg in metal-poor AGB models.
In addition, we enhance the other ${\alpha}$~elements by $+0.4$ dex.

We assume mixing begins on the part of the RGB where the H~shell
 burns through the hydrogen discontinuity left behind after the first
 dredge-up, in accordance with the assumption that 
 large ${\mu}$ barriers can prevent mixing at earlier epochs (SM79).
This point along the RGB corresponds to the well-known ``bump'' in the
 luminosity function \citep{BUMP}.
Supporting this choice are the theoretical mixing models by
 \citet{Charbonnel94,Charbonnel95} that also assumed
 mixing begins at this point and reproduced the observed
 variations of the {\cratio}, $^{7}$Li and the $^{12}$C/$^{14}$N ratio
 in both open and globular clusters.
In addition, the observations of \citet{BDG79}, \citet{NBS81}, \citet{GB91},
 \citet{FG99} and \citet{CGSB99} also support this choice for the onset of
 mixing.
We do point out that not all observations support this choice as
 the onset mixing \citep{Carbon1982,TCLSK83,LKCF86,BBHD90}, however,
 the exact start of deep mixing will little affect our final results.

The timestep for nuclear processing fixes the timestep for mixing and
 is controlled by setting a limit on how fast any element above a minimum
 abundance threshold may vary.
Since this timestep is much shorter than the time difference between the models
 used in CSB98, new models were interpolated along the RGB until
 the He flash was encountered.
The free parameters in our code are ${\Delta}X$, the mixing depth defined
 by a change in the H-mass fraction, $X$, within the H~shell, and ${\eta}$, 
 Reimers' mass-loss parameter \citep{Reimers75}.
The algorithm was run with various combinations of mixing depths and
 mass-loss rates for a stellar evolution sequence having [Fe/H]$~=~-1.67$.

Figure 13 shows the [Al/Fe] values derived from our algorithm 
 as a function of absolute magnitude, parameterized by various mixing 
 depths and mass-loss rates.
The absolute magnitude scale was derived from the bolometric luminosity, $L$,
 and {\teff} provided by the models.
We first converted log ($L/{\rm L}_{\odot}$) into the bolometric magnitude,
 M$_{bol}$, using the sun as a reference with a value of 4.75 for 
 M$_{bol\odot}$.
Next, we used the following relationship between gravity, $g$, mass, $M$,
 {\teff} and $L$ to obtain log $g$ for each model:
\begin{equation}
   {\rm log}~g~=~{\rm log}~(M/{\rm M}_{\odot})~+~4{\rm log}~T_{\rm eff}~
   -~{\rm log}~(L/{\rm L}_{\odot})~-~10.61028,
\end{equation}
 where a value of T$_{{\rm eff}{\odot}}~=~5780~{\rm K}$ is assumed.
Mass loss is implicitly accounted for via the first term on the right hand 
 side of equation 1.
Using a 12 Gyr isochrone with [Fe/H]~=~$-$1.67, also constructed by Drs.
 VandenBerg and Bell,
 we interpolated according to log $g$ to find a bolometric correction then
 converted M$_{bol}$ into M$_{V}$.
We show the results for ${\Delta}X$ = 0.05, 0.10, 0.15 and 0.20, and for
 ${\eta}$ = 0.0, 0.2, 0.4 and 0.6 at each ${\Delta}X$, as described
 in the figure caption.

\begin{figure}
\plotone{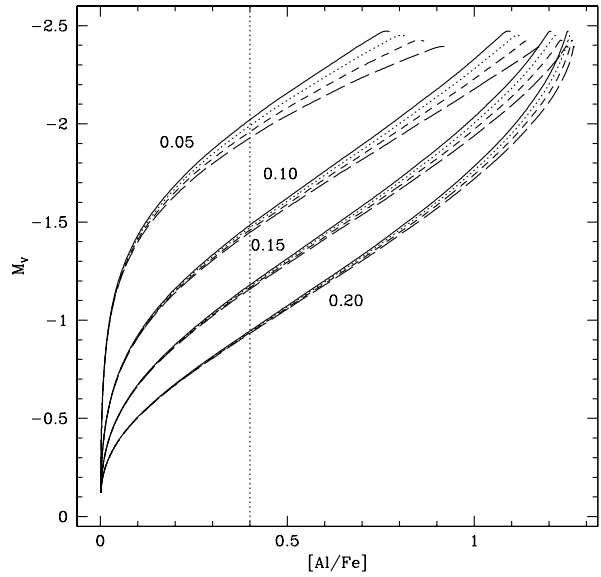}
\caption[Cavallo.fig13.eps]{The predicted variation in [Al/Fe] with absolute
 magnitude for different mixing depths and mass-loss parameters.  The
 groups of lines are labeled by the mixing depth, ${\Delta}X$.
 The long-dashed, short-dashed, dotted and solid lines are for Reimers' (1975)
 ${\eta}$ = 0.6, 0.4, 0.2 and 0.0, respectively.
 The vertical dotted lines represents the low end of ``high'' [Al/Fe] values 
 in M~13.}
\end{figure}

According to Figure 13, the dominant parameter for determining [Al/Fe] at
 the RGB tip is the mixing depth.
Mass loss plays a secondary role for deeply mixed models but is more
 important for less deeply mixed models.
The importance of ${\eta}$  is controlled by a competition between the timescale
 for mass loss and the timescale for converting magnesium into aluminum:
 for a given sequence, the mass-loss timescale is fixed by ${\eta}$, so
 that with deeper mixing, the Mg-burning timescale decreases, bringing
 the two closer together and limiting the influence of mass loss.
Thus, in the limit of instantaneous mixing, the distribution of [Al/Fe] near
 the tip of the RGB is due primarily to variations in the mixing depth; 
 although, for the deeply mixed sequences, the value of [Al/Fe] at the tip
 begins to saturate.

One factor that depends strongly on the mixing depth is the earliest point
 along the RGB where mixing-induced aluminum variations can occur.
In Figure 13 we draw a dashed vertical line at [Al/Fe]~=~$+0.4$ to indicate
 where the aluminum enhancements cross the into the ``high'' aluminum 
 distribution for M~13 giants.
Our results show that for the mixing depths shown in Figure 13, 
 large aluminum enhancements should appear along the brightest
 ${\sim}$~1.5 to 0.5 magnitudes of the RGB.
A change in ${\Delta}X$ from 0.05 to 0.20 results in a one magnitude 
 difference in where the aluminum abundance rises on the RGB.
We apply these estimates to the M~13 sample by binning the data according to
 the magnitudes at which the various mixing depths predict the aluminum
 abundances will cross the [Al/Fe]~=~$+0.4$ threshold.
The results are shown in Table 8, where the first two columns describe the
 models and the next five discuss the data.  
The first column gives the mixing depth and the second column describes how
 far down the RGB the models predict a star will cross into the ``high''
 aluminum group for that mixing depth, while
 the third column gives the fraction of stars in the bin (out of 96).
The fourth and fifth columns give the $P$-value and the Al ratio for stars in
 the bin, respectively, while final two columns give the $P$-value and the 
 Al ratio for the remaining stars outside the bin.
\footnote{It must be emphasized again that the KMM algorithm is only accurate
 for sample sizes greater than 50; nonetheless, we include the $P$-values for
 the sake of completeness.}.

If the assumption that mixing is the cause of the aluminum enhancements in
 the bright giants of M~13 holds, then it is apparent from Table 8
 that it must turn on somewhere during the brightest magnitude of the RGB,
 as the Al ratio changes from 67:33 to 88:12 as one approaches the RGB tip.
This signifies that at least 21\% of the giants are experiencing deep mixing.
We call this a lower limit because our technique for measuring aluminum 
 enhancements cannot detect mixing in stars with initially high [Al/Fe] values.
Since two-thirds of the giants in M~13 appear to have high aluminum
 abundances before mixing takes effect, we could be missing a substantial
 number of stars undergoing deep mixing.
If the same relative number of giants with initially high aluminum abundances
 undergo deep mixing as the relative number of giants with initially low
 aluminum abundances, the percentage of all stars undergoing deep mixing jumps
 to 63\%.

\begin{deluxetable}{ccccccc}
\tablewidth{0pc}
\tablenum{8}
\tablecaption{Ratio of Al-enhanced to Al-normal Giants in M~13}
\tablehead{
 \colhead{${\Delta}X$}   &
 \colhead{${\Delta}M_V$} &
 \colhead{\% Stars}      &
 \colhead{$P$-value}     &
 \colhead{Al Ratio}      &
 \colhead{$P$-value}     &
 \colhead{Al Ratio}      \\
 \colhead{}              &
 \colhead{from RGB tip}  &
 \colhead{IN}            &
 \colhead{IN}            &
 \colhead{IN}            &
 \colhead{OUT}           &
 \colhead{OUT}
}
%
\startdata
0.05 & 0.50 & 18 & 0.005 & 7.5${\pm}$5.6 & 0.000 & 1.7${\pm}$0.2 \\
0.10 & 1.00 & 35 & 0.000 & 2.8${\pm}$1.1 & 0.007 & 1.7${\pm}$0.5 \\
0.15 & 1.30 & 44 & 0.001 & 2.5${\pm}$0.9 & 0.005 & 1.8${\pm}$0.5 \\
0.20 & 1.50 & 46 & 0.001 & 2.4${\pm}$0.8 & 0.006 & 1.9${\pm}$0.6 \\
\enddata
\end{deluxetable}

\subsection{Deep Mixing, the Blue-Tail Parameter and the Signatures of
 AGB Pollution}

Deep mixing in red giants might have an effect on their future evolution.
For example, \citet{CG96} noticed a relationship between the HB
 morphology and the amount of depletion of oxygen in RGB stars.
This does not necessarily imply that oxygen is a second parameter,
 but rather that whatever is responsible for the oxygen depletions might
 also be causing the blueward shift in the HB.
One such mechanism that can do both is rotation,
 which has several effects: 1) it can extend the life of a red giant,
 causing it to lose more mass and ultimately end up on the blue HB,
 2) it can drive meridional circulation currents, which can deplete the
 oxygen, and 3) if fast enough, it can cause the circulation
 currents to penetrate the H~shell and bring helium to the surface.
This extra helium causes RGB stars to evolve to the blue HB at brighter
 luminosities than their unmixed counterparts, mimicking the second-parameter
 effect and reproducing the upward slope of the HB with decreasing color
 \citep{SC98} that is observed in some metal-rich clusters
 \citep{PIOTTO97,RICH97}.

In this section, we examine the suggestion by \citet{Ferr98} that a so-called
 ``blue-tail second parameter'' (BTP) exists in M~13.
Such a parameter differs from the more commonly sought after second parameter
 in the sense that the latter typically deals with difference in HB
 morphology on the flat part of the HB, whereas the former describes
 how clusters like M~13, M~80 and NGC~6752 develop extended blue tails.
We attempt to discover whether or not deep mixing can be a/the BTP by
 adding extra helium into the atmosphere of RGB stars.
Unfortunately, helium cannot be measured spectroscopically in cool giants;
 however, as shown in Figure 8, aluminum is made from magnesium inside
 the H~shell where helium is being produced, so that the mixing of helium
 is accompanied by the mixing of aluminum; i.e., aluminum can be a good
 tracer of helium mixing.
We conclude that if deep mixing is a BTP and if aluminum
 traces helium mixing, then there should exist a correlation
 between the Al ratio and the HB morphology. 
To describe the HB morphology quantitatively, we suggest using at the ratio 
 of blue to red HB stars (hereafter, the ``HB ratio''), which, of course,
 require definitions of their own.
Perhaps a solution can be found in the corollary assumption that if
 cluster giants do not mix, then the cluster should not have an
 extended blue tail on the HB.
Therefore, by assumption, a cluster like M~3, whose giants appear not to
 experience deep mixing, defines the ``red'' HB, so that for clusters
 like M~13, any star on the HB that is hotter than the M~3 HB is defined
 as ``blue,'' provided, of course, that the clusters are similar in all
 other ways (e.g., metallicity, age, environment, etc.).

To make the comparison between M~3 and M~13, we obtained high-quality Hubble
 Space Telescope photometric data for both clusters from Dr. F. Ferraro and
 shifted the M~3 HB by ${\delta}V~=~-0.6$ and ${\delta}(U-V)~=~-0.03$
 to align it with that of M~13, as done by \citet{Ferr98}.
We then plotted histograms of the distributions of colors along the HB's for
 each cluster, as shown in Figure 14, and compared the HB ratio in M~13 with
 its Al ratio.
We define stars with $U-V~<~-0.3$ as being blue, which gives a HB ratio
 of 58:42.
We note that while this choice of color coincides with the apparent gap 
 in Figure 14, it is chosen because this is where the M~3 distribution drops 
 off and not because of the appearance of bimodality in the M~13 distribution.
To estimate an error in the HB ratio, we fit a Gaussian to the M~3
 distribution to determine the standard deviation, ${\sigma}$,
 and call blue all M~13 stars hotter than the mean minus $~3{\sigma}$ in M~3,
 resulting in an HB ratio of 74:26.
Compared with the 21\% to 63\% deeply mixed stars on the RGB,
 the 58\% to 74\%  blue HB stars is consistent with deep mixing as the BTP;
 although, the uncertainties in the number of RGB stars that have undergone
 deep mixing makes the results less robust than desired.

\begin{figure}
\plotone{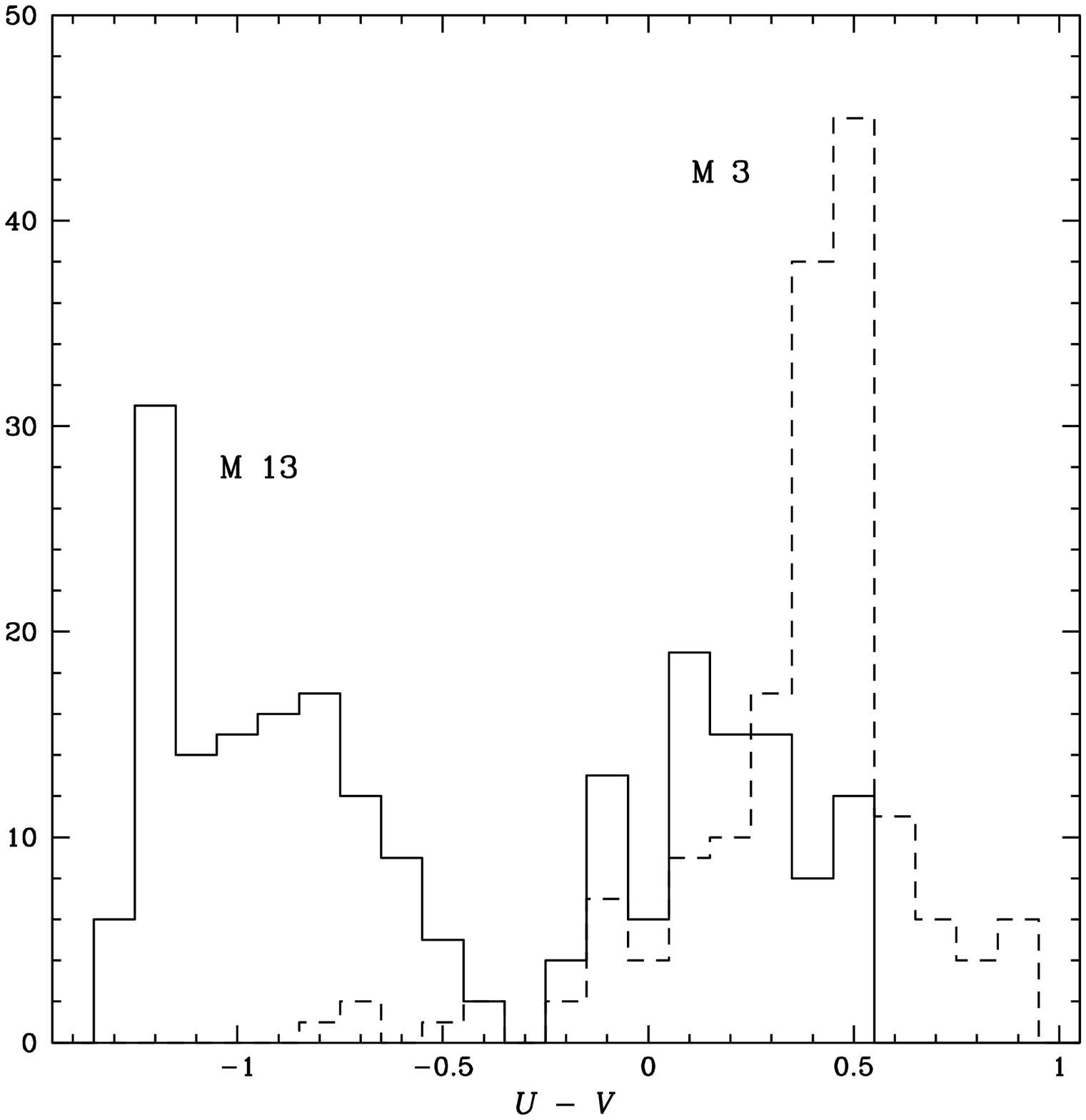}
\caption[Cavallo.fig14.eps]{Histograms of the HB distributions in M~3 (dashed line)
 and M~13 (solid line).  The photometry is from \citet{Ferr98} and the
 M~3 data have been shifted by ${\delta}V~=~-0.6$ and ${\delta}(U-V)~=~-0.03$.}
\end{figure}

It would be helpful if we could discriminate between stars that have undergone
 deep mixing and those that have been polluted.
One way to do this might be by using the $s$-process abundances, which are
 also created in intermediate-mass AGB stars (IMS) with aluminum, albeit at 
 different locations within the stars \citep{DDNW98,Lattanzio99,BS99}. 
Stars above $M~{\gtrsim}4~{\rm M}_{\odot}$ experience HBB, which, as discussed
 above, result in the production of $^{27}$Al from magnesium.
These same stars also create a neutron exposure during the thermal pulses
 in the He-burning shell that favors the production of the Sr-Y-Zr peak
 elements and $^{25}$Mg, all through the $^{22}$Ne$({\alpha},n)^{25}$Mg
 reaction \citep{GBLTS99}.
If the $n$-rich material is mixed with the HBB by-products and ejected
 into the cluster medium, then one should trace the other in the
 polluted stars.
Specifically, we expect stars with high $s$-process abundances to also have 
 high [Al/Fe] values, but not vice-versa.
The key is to choose the best {\spr} elements that trace the Al-rich
 IMS ejecta.
We suggest using zirconium since many lines are available in the
 optical spectrum and it's abundance easily computed \citep{CN2000}.
Conversely, elements near the barium peak would not be good choices
 to represent the {\spr}/HBB enhancement from IMS since they are produced in
 low-mass AGB stars ($M~{\sim}2~{\rm M}_{\odot}$).

\subsection{On the Overproduction of Sodium}

Recently, \citet{CDW99} pointed out that mixing into the H~shell to enrich
 the stellar atmosphere in aluminum and helium would result in an
 overproduction of sodium by ${\gtrsim}~0.3$ dex (see their Figure 3) 
 relative to the M~13 data, essentially precluding deep mixing.
This constraint keeps the change in the atmospheric helium abundance
 to less than 0.06, much less than the 0.20 found by \citet{AVS97a,AVS97b} 
 that is needed to account for the most extended HB's seen in clusters 
 like M~13.
We submit that the solution to this discrepancy might be found in the initial 
 abundances; i.e., primordial contamination from the IMS plays a role.
Our algorithm shows that the overproduction of sodium is avoided, even with
 deep mixing, if the initial $^{22}$Ne abundance were depleted as a result
 of $^{22}$Ne$({\alpha},n)^{25}$Mg reaction in the IMS, which also 
 enhances the initial $^{25}$Mg, as needed to produce the large aluminum
 enhancements in the RGB stars.
The calculations of \citet{GBLTS99} indicate that the $^{22}$Ne abundance
 is depleted by approximately 30\% during the thermal pulses in
 IMS, but it is not clear how HBB and interpulse burning affects
 the net $^{22}$Ne abundance.
One would assume that HBB would deplete the $^{22}$Ne reserves in the
 convective envelope and produce $^{23}$Na as is done on the RGB.
In contradistinction, the AGB yields calculated by \citet{DDNW98} actually 
 enhance the net $^{22}$Ne abundance from a series of ${\alpha}$ captures 
 on $^{14}$N.
Clearly, a more complete and detailed look into the yields 
 of all abundances from primordial AGB stars is necessary to determine a
 more translucent picture of how pollution plays a role on the RGB.

Our hypothesis is consistent with the sodium data, which raise two
 important questions: 1) why does the Na ratio increase only at the same
 magnitudes as the Al ratio when sodium is very easily produced from
 $^{22}$Ne above the H~shell at luminosities far below the RGB tip, and
 2) why does the sodium abundance vary without oxygen abundance variations
 for ``oxygen-normal'' giants (see, e.g., KSLP).
The answer to both these questions might be found on the AGB: if
 $^{22}$Ne is depleted to build up $^{25}$Mg during the He-shell flashes
 and $^{23}$Na during HBB, then sodium will not be produced at lower
 magnitudes on the RGB, but will be made at brighter magnitudes
 from $^{20}$Ne with deep mixing.
The extra sodium produced during HBB could be distributed locally within
 the cluster creating the [Na/Fe]-rich stars that are independent of their
 oxygen abundances.
Although oxygen is likely depleted during HBB, this is unlikely to create
 an oxygen-poor RGB atmosphere since it is easier to enhance elements
 in an atmosphere than to deplete them by pollution.
A primordial pollution scenario is consistent with the data that show
 aluminum and sodium enhancements on the lower RGB and, for some clusters, 
 on the main sequence and subgiant branch, and can help prevent the 
 overproduction of sodium during deep mixing.

\section{CONCLUSIONS}

Before we discuss our final conclusions, we first remind the reader of the 
 number of assumptions that have gone into our analysis.
First, there are errors associated with the abundance determinations that
 we tried to characterize by allowing for significant variations in the
 model atmosphere parameters, which contribute the most to the uncertainty
 in the analysis.
Second, the inclusion of the WIYN data into our analysis comes at a price:
 the data have poor signal-to-noise ratios, come from only one line and
 require the assumption that, for some stars with indeterminate line strengths,
 the [Al/Fe] value is ``low.''
Third, despite that fact that this is the largest compilation of [Al/Fe] 
 and [Na/Fe] values in one globular cluster to be analyzed in a single
 paper, the data are still subject to small number effects, particularly
 at the RGB tip.
Unfortunately, there are only so many tip stars that can be spectroscopically
 measured from the ground, leaving this problem difficult to solve.
We suggest the best way to handle the small numbers is to expand this
 analysis to other clusters for a broad comparative study.
Fourth, the models have many assumptions in them: we assume that canonical
 evolution holds and add in our mixing algorithm after the fact, we assume
 that mixing is instantaneous, we assume that the abundances are distributed
 as per \citet{DDNW98}, we assume that our reaction rates are accurate, and
 we assume that we adequately searched the parameter space allowed by
 the uncertainties in the initial abundances, nuclear reaction rates and
 mass-loss rates.
Fifth, we assume that no other second parameter affects the relationship
 between the M~3 and M~13 HB morphologies.
Sixth, we make no attempt to correct for blending of the AGB with the RGB
 when performing our analysis.
Approximately 20\% of the red giants above the point where giant branches
 merge are supposedly AGB interlopers based on comparative lifetimes:
 the problem is to determine which ones are really AGB stars.
This might not be as much of a problem for the M~13 sample, however, since
 blue HB stars tend to evolve away from the AGB.
The best workaround for this problem is also an extension of our analysis to
 other clusters to look for consistent trends despite this, and the other,
 uncertainties.

The importance of having accurate nuclear reaction rates cannot be overstated.
This is particularly true when using aluminum as a diagnostic of deep mixing.
If we were to vary, for example, the $^{26}$Mg proton-capture rate to its
 upper limit in range of $T_9~=~0.05-0.06$, the production of Al can move
 outside the H shell, although, just barely.
Depending on the initial abundances of $^{25}$Mg and $^{26}$Mg, this might
 be able to account for the full enhancements of aluminum that we observed.
In addition, according to the NACRE compilation, the rate for
 $^{26}$Al$^{g}(p,{\gamma})^{27}$Si is uncertain by as much as three orders of
 magnitude in the same temperature range.
Increasing these rates might help solve the problem presented by \citet{CDW99}
 who show that, if mixing occurs below the top of the H shell, sodium is 
 overproduced due to the extra enhancement from $^{20}$Ne in the NeNa cycle,
 a result we confirm with our instantaneous mixing algorithm.
If the $^{26}$Mg proton-capture rate is near its upper limit, then deep
 mixing is not required to produce the observed aluminum abundances and
 sodium is not over enhanced compared to the observations.

Our general results for the M~3 and M~13 abundances obtained in this
 work show the usual trends in the proton-capture, ${\alpha}$ and
 iron-peak elements:
 the sodium and aluminum abundances are anticorrelated with oxygen, the
 ${\alpha}$ elements are enhanced by approximately 0.3 dex and the
 iron-peak elements remain constant.

Our analysis shows that the variation in both the [Al/Fe] and [Na/Fe] ratios
 are consistent with deep mixing occurring on the RGB in M~13 and not in M~3.
The aluminum and sodium data are correlated for the M~13 giants;
 although, the Al ratio is probably a better indicator of deep mixing
 since it is more easily separated into ``high'' and ``low'' groups.
We would not expect such a similar tight correlation between aluminum and
 sodium in the M~3 giants since sodium can be enhanced without increasing
 the aluminum abundance if the mixing currents do not penetrate the H shell,
 as seems indicated in M~3 from the low Al ratio.
However, some semblance of a correlation between aluminum and sodium might be
 set up by primordial effects in this cluster.
In addition, the Na ratio increases near the same magnitude as the Al ratio,
 which is contrary to the previous predictions that sodium should be
 enhanced further down the RGB from $^{22}$Ne (CSB98).
Our models show that this would be expected if the $^{22}$Ne were depleted
 in primordial intermediate-mass AGB stars.

When comparing the Al ratio with the HB ratio, it seems that the
 assumption of deep mixing as a blue-tail parameter is self-consistent;
 however, the large range allowed in the actual number of mixed RGB stars
 and the empirical definitions of ``blue'' and ``red'' HB stars do
 not constrain the results enough to be firmly conclusive.
Again we suggest that a similar analysis as the one presented here be extended
 to other clusters to determine the Al ratio as a function of $V$ and
 to compare this with the HB ratio.
If the Al ratio at the RGB tip can be shown to be a predictor of the 
 HB ratio, then helium mixing would certainly be given greater credence
 as a blue-tail second parameter, supplanting the oft-assumed cluster age
 differences that have been shown to fail for this classical pair of clusters.
In particular, we suggest the study of metal-rich clusters to see if the
 aluminum distribution is bimodal, and if it is, if the Al ratio varies.
According to our models, it should not vary since aluminum cannot be produced
 in metal-rich cluster giants on the same scale as it can in the intermediate
 metallicity and metal-poor giants.
In addition, we suggest further examination of the sodium abundance in
 clusters to search for similar behavior as in M~13.
Also, we suggest a more extensive comparison of the {\spr} abundances with
 the aluminum data as a test of primordial contamination.

Finally, we conclude that the problem of abundance anomalies in globular
 cluster red giants requires detailed study of the abundance yields
 from primordial AGB stars as well as an in-depth and complete study
 of the hydrodynamical evolution of rotating RGB stars.
In the meantime, aluminum, and to a lesser extent, sodium, give the
 best diagnostics of deep mixing during the evolution up the RGB and the 
 {\spr} elements near the Sr-Y-Zr peak are the best tracers of AGB pollution
 from IMS.

\acknowledgments
The authors expressly acknowledge and thank Caty Pilachowski for allowing
 us to use her data and for her valuable help in analyzing them.
We also extend our gratitude to Mike Briley, who supplied us 
 with the unpublished data for the star M~3 AA and who acted as our
 referee with many thoughtful and valuable comments.
We thank Bob Kraft and Chris Sneden, who sent us their original Lick spectra
 for the stars in common between our two studies, and to Michael Bolte, who
 provided us with finding charts for the star MB~4.
We express our gratitude to Francesco Ferraro for making 
 his recent HB photometry for M~13 and M~3 available to us.
We thank Carrie Rowland for giving us updated data for
 the $^{26}$Mg proton capture rates and we look forward to her
 experimental results regarding this reaction.
Our thanks are given to Keith Ashman for supplying us with a FORTRAN version
 of his KMM algorithm.
In addition, we are indebted to Roger Bell for providing us with
 his latest isochrones describing color-{\teff} relations.
We also thank Allen Sweigart for his many invaluable discussions throughout the
 development of this project.
We wish to thank Daryl Wilmarth for his aid in gathering and
 reducing the data and Frank Hill for providing us with the Kitt Peak
 solar spectrum.
We likewise acknowledge Jennifer Johnson and Peter Stetson for making 
 their separate photometric data sets available to us.
N. M. N. acknowledges travel support from the Astronomy Department at UMD and
R. M. C acknowledges KPNO for travel support while he was a visiting graduate
 student.
This work was performed while R. M. C. held a National Research Council-GSFC
 Research Associateship.
Funding for publication was provided in part by a Small Research Grant from
 the American Astronomical Society.

\appendix
\section{INSTANTANEOUS MIXING ALGORITHM}

This derivation of instantaneous mixing begins with the form of the nuclear
 reaction equation that involves the proton-capture reactions and 
 ${\beta}$-decays; although, it is easily extended to other rates and is
 applied in its most general form in our code:
\begin{equation}
 \frac{{\rm d}n_{i}}{\rm dt}~=~\sum_{j} ({\pm}n_{j}n_{\rm H}<{\sigma}v>_{j}~
                               {\pm}~n_{j}\frac{{\ln}2}{{\tau}_{j}}),
\end{equation}
 where $n_{i}$ is the number of nuclei of type $i$ cm$^{-3}$ that are
 being produced or destroyed, $n_{j}$ is the number cm$^{-3}$ of nuclei that
 produce ($+$ sign) or, when $j~=~i$, destroy ($-$ sign) nuclei of type ${i}$,
 $n_{\rm H}$ is the number cm$^{-3}$ of protons, $<{\sigma}v>_{j}$ is the 
 velocity-averaged cross section of the proton-capture reaction and
 ${\tau}_{j}$ is the mean lifetime of radioactive isotopes that destroy
 or produce element $i$.

If we integrate equation A1 over a mass interval from some mixing depth, 
 $M_{\rm d}$, to the surface,
 $M$, and substitute the molar fraction $Y~=~n/{\rho}{\rm N_A}$, where 
 ${\rho}$ is the density and N$_{\rm A}$ is Avogadro's number, so that
 $Y_{\rm H}~=~X$, the hydrogen-mass fraction, we get
\begin{equation}
 \int_{M_{\rm d}}^{M} \frac{{\rm d}Y_{i}}{\rm dt} {\rm d}M_{r}~=~
 \sum_{j} 
     ({\pm} \int_{M_{\rm d}}^{M} 
             Y_{j}X{\rm N_A}<{\sigma}v>_{j}{\rho}{\rm d}M_{r}~
     {\pm}~\frac{{\ln}2}{{\tau}_{j}} \int_{M_{\rm d}}^{M} Y_{j}{\rm d}M_{r}),
\end{equation}
 which is equivalent to spreading the nuclearly processed material over the
 whole mixing zone.
Mass loss can be accounted for my modifying the total integrated mass by
 some mass-loss recipe such as given by \citet{Reimers75}.

Now assuming that the mixing is instantaneous so that d$Y_{i}$/dt, $Y_{j}$ 
 and $X$ vary little over the whole mixing zone, we can rewrite equation
 A2 as
\begin{equation}
 \frac{{\rm d}Y_{i}}{\rm dt} \int_{M_{\rm d}}^{M}{\rm d}M_{r}
                            ~=~
 \sum_{j} 
      ({\pm} Y_{j}X
      \int_{M_{\rm d}}^{M} 
           {\rm N_A}<{\sigma}v>_{j}{\rho}{\rm d}M_{r}
                             ~{\pm}~
       \frac{{\ln}2}{{\tau}_{j}} Y_{j} \int_{M_{\rm d}}^{M}{\rm d}M_{r}).
\end{equation}

The first integral on the right hand side is just the mass-averaged 
 reaction rate that can be substituted as an effective reaction rate, 
 $<{\sigma}v>_{j}^{\rm eff}$, while the other two integrals in equation
 A3 are just the total mixed mass, $M_{\rm mix}$.
Thus, equation A3 can now be written 
\begin{equation}
 \frac{{\rm d}Y_{i}}{\rm dt}
         ~=~
 \frac{1}{M_{\rm mix}} 
 \sum_{j} 
    ({\pm} Y_{j}X<{\sigma}v>_{j}^{\rm eff}
                 ~{\pm}~
    \frac{{\ln}2}{{\tau}_{j}} Y_{j}),
\end{equation}
 the final form of equation A1 under the assumptions of instantaneous mixing.

The implementation of this equation in our nuclear reaction network
 is straightforward.
We average the reaction rates together by weighting the reaction
 rate determined at each mesh point by the mass contained between that
 mesh point and the one below it and summing over all the mass intervals.
The temperature and density for calculating the reaction rate for each
 mass interval are taken at the top mesh point (towards the surface).
Since the spacing between mesh points becomes closer as the temperature
 profile steepens, the differences between the temperature and density at
 the top and bottom of the mass intervals has negligible influence on the
 effective reaction rates.
The effective rates are then applied to the initial abundances and
 integrated (``mixed'') over some determined mass interval where the
 mixing depth and mass-loss rate are the free parameters chosen by the user.
The burning timescale is controlled by limiting how much the fastest 
 changing isotope with some chosen minimum abundance can vary in a
 single timestep.
New models are interpolated from a sequence at timesteps
 according to this nuclear burning timescale.
Each new model contains the output abundances derived from the previous model
 for its input abundances.
The mixing algorithm can begin anywhere on the RGB and proceeds until the
 helium flash is encountered at the tip.
The code outputs the new abundances, the mass-averaged reaction rates and
 information regarding the position of the model on the theoretical RGB.

\begin{deluxetable}{lllllllllllll}
\tabletypesize{\footnotesize}
\tablenum{2}
\tablewidth{0pc}
\tablecaption{Measured Equivalent Widths (m{\AA})}
\tablecolumns{13}
\tablehead{
 \multicolumn{3}{c}{}            & 
 \multicolumn{6}{c}{M~3 Giants}  &  
 \colhead{}                      & 
 \multicolumn{3}{c}{M~13 Giants} \\
 \cline{4-9}  \cline{11-13}
 \colhead{Wavelength}  & 
 \colhead{E. P.}       & 
 \colhead{log ($gf$)}  & 
 \colhead{AA}          & 
 \colhead{vZ~297}      & 
 \colhead{III-28}      & 
 \colhead{vZ~1000}     & 
 \colhead{vZ~1127}     & 
 \colhead{MB 4}        & 
 \colhead{}            & 
 \colhead{L~262}       & 
 \colhead{L~324}       & 
 \colhead{III-56}      \\
 \colhead{(\AA)}       & 
 \colhead{(eV)}        & 
 \multicolumn{11}{c}{}
}
\startdata
\cutinhead{Na I}
6154.23 &  2.10 &  $-1.66$    &  33.5    &  20.7    &  \nodata &  \nodata &  \nodata &  34.1    & &  20.0    &  34.5    &  29.1    \\*
6160.75 &  2.10 &  $-1.32$    &  51.7    &  29.8    &  \nodata &  33.2    &  14.9    &  52.1    & &  38.6    &  53.1    &  48.9    \\*
\cutinhead{Mg I}
5711.09 &  4.35 &  $-1.58$    &  110.7   &  93.2    &  73.9    &  93.8    &  83.5    &  96.4    & &  91.5    &  87.2    &  100.4   \\*
8717.83 &  5.93 &  $-0.96$    &  \nodata &  \nodata &  \nodata &  18.8    &  15.7    &  21.5    & &  11.9    &  10.4    &  18.4    \\*
8736.02 &  5.95 &  $-0.04$    &  \nodata &  59.1    &  50.7    &  74.0    &  54.1    &  68.2    & &  51.6    &  45.5    &  52.7    \\*
\cutinhead{Al I}
6696.03 &  3.14 &  $-1.54$    &  69.0    &  52.4    &  \nodata &  51.1    &  13.2    &  58.8    & &  41.1    &  71.1    &  52.8    \\*
6698.67 &  3.14 &  $-1.91$    &  40.1    &  24.6    &  \nodata &  25.9    &   5.1    &  34.5    & &  18.2    &  36.9    &  23.7    \\*
7835.31 &  4.02 &  $-0.72$    &  36.0    &  28.3    &  \nodata &  26.3    &  \nodata &  31.3    & &  18.3    &  38.7    &  23.4    \\*
7836.13 &  4.02 &  $-0.63$    &  50.9    &  36.3    &  \nodata &  30.2    &  \nodata &  45.8    & &  24.2    &  47.3    &  32.7    \\*
8772.86 &  4.02 &  $-0.35$    &  63.1    &  60.1    &  \nodata &  59.8    &  \nodata &  64.0    & &  38.0    &  80.1    &  51.2    \\*
8773.90 &  4.02 &  $-0.15$    &  91.3    &  83.8    &  \nodata &  81.0    &  \nodata &  98.1    & &  62.1    &  98.8    &  62.2    \\*
\cutinhead{Si I}
6142.48 &  5.62 &  $-1.53$    &  10.2    &  \nodata &  \nodata &  \nodata &  \nodata &  6.1     & &  11.4    &  9.3     &  6.2     \\*
6145.02 &  5.61 &  $-1.41$    &  \nodata &  \nodata &  \nodata &  \nodata &  \nodata &  \nodata & &  \nodata &  10.1    &  9.9     \\*
6243.82 &  5.61 &  $-1.32$    &  17.1    &  14.8    &  11.7    &  16.7    &  14.0    &  15.4    & &  15.3    &  11.9    &  \nodata \\*
6721.85 &  5.86 &  $-1.16$    &  \nodata &  \nodata &  9.0     &  14.0    &  12.3    &  \nodata & &  14.5    &  9.3     &  13.4    \\*
7405.77 &  5.61 &  $-0.57$    &  \nodata &  60.2    &  50.0    &  53.2    &  45.6    &  42.0    & &  \nodata &  55.0    &  41.7    \\*
7415.95 &  5.68 &  $-0.77$    &  \nodata &  44.7    &  31.5    &  43.6    &  33.4    &  28.6    & &  45.2    &  36.8    &  34.6    \\*
8648.47 &  6.21 &  $-0.10$    &  47.9    &  \nodata &  \nodata &  \nodata &  \nodata &  \nodata & &  \nodata &  \nodata &  \nodata \\*
8742.45 &  5.87 &  $-0.41$    &  \nodata &  41.6    &  45.3    &  45.0    &  38.0    &  42.0    & &  \nodata &   41.1   &  36.3    \\*
8752.01 &  5.87 &  $-0.42$    &  \nodata &  56.2    &  36.2    &  47.4    &  47.4    &  35.8    & &  54.1    &  47.2    &  37.9    \\*
\cutinhead{Ca I}
6161.30 &  2.52 &  $-1.22$    &  87.7    &  88.0    &  70.5    &  78.2    &  78.5    &  91.8    & &  69.9    &  78.4    &  85.8    \\*
6166.44 &  2.52 &  $-1.04$    &  95.5    &  90.1    &  64.6    &  \nodata &  74.6    &  96.2    & &  74.5    &  83.3    &  78.1    \\*
6169.04 &  2.52 &  $-0.67$    &  115.9   &  112.5   &  \nodata &  108.0   &  96.7    &  113.8   & &  96.3    &  115.7   &  113.1   \\*
6439.08 &  2.53 &  {\phs}0.26 &  197.3   &  182.1   &  166.7   &  176.5   &  169.3   &  182.9   & &  166.7   &  196.8   &  185.8   \\*
6455.60 &  2.52 &  $-1.34$    &  78.4    &  71.6    &  52.5    &  62.9    &  61.2    &  74.7    & &  57.6    &  64.8    &  68.5    \\*
6471.66 &  2.53 &  $-0.67$    &  123.1   &  113.5   &  99.2    &  112.9   &  102.2   &  123.8   & &  108.3   &  131.0   &  120.3   \\*
6493.78 &  2.52 &  $-0.22$    &  162.3   &  153.6   &  128.6   &  145.9   &  132.0   &  159.7   & &  129.4   &  163.6   &  152.7   \\*
6499.65 &  2.52 &  $-0.79$    &  119.7   &  103.0   &  96.4    &  105.0   &  91.8    &  109.7   & &  89.3    &  107.8   &  104.4   \\*
\cutinhead{Ti I}
5866.45 &  1.07 &  $-0.82$    &  154.8   &  137.2   &  110.1   &  115.4   &  113.9   &  144.5   & &  112.8   &  156.0   &  131.9   \\*
6064.63 &  1.05 &  $-1.99$    &  72.4    &  58.0    &  36.3    &  42.9    &  38.0    &  \nodata & &  40.0    &  65.3    &  59.1    \\*
6126.22 &  1.07 &  $-1.43$    &  113.9   &  95.4    &  75.3    &  77.6    &  84.8    &  108.9   & &  81.2    &  104.0   &  93.2    \\*
6258.10 &  1.44 &  $-0.37$    &  149.1   &  140.9   &  99.9    &  115.8   &  115.3   &  156.6   & &  109.9   &  162.7   &  136.4   \\*
6261.10 &  1.43 &  $-0.46$    &  159.1   &  138.4   &  96.9    &  119.6   &  114.4   &  147.4   & &  108.4   &  154.9   &  136.2   \\*
6336.10 &  1.44 &  $-1.80$    &  36.6    &  35.4    &  17.3    &  28.1    &  20.8    &  41.9    & &  20.7    &  31.5    &  22.9    \\*
6497.68 &  1.44 &  $-2.10$    &  29.5    &  13.0    &  13.5    &  10.8    &  14.0    &  28.6    & &  11.1    &  19.8    &  14.3    \\*
6508.12 &  1.43 &  $-2.18$    &  28.9    &  14.1    &  10.2    &  14.5    &  11.7    &  23.7    & &  12.5    &  20.6    &  14.8    \\*
\cutinhead{Ti II}
6219.94 &  2.06 &  $-3.14$    &  \nodata &  \nodata &  9.7     &  9.2     &  \nodata &  12.7    & &  14.3    &  12.7    &  16.7    \\*
6491.56 &  2.06 &  $-2.07$    &  75.1    &  63.5    &  63.0    &  65.9    &  66.4    &  61.7    & &  66.3    &  72.0    &  71.8    \\*
\cutinhead{V I}
5670.85 &  1.08 &  $-0.47$    &  100.8   &  78.2    &  50.6    &  67.9    &  49.2    &  85.1    & &  60.8    &  81.5    &  70.8    \\*
5727.05 &  1.08 &  {\phs}0.02 &  142.4   &  120.3   &  80.2    &  105.3   &  99.9    &  \nodata & &  98.3    &  128.4   &  117.3   \\*
5727.65 &  1.05 &  $-0.92$    &  68.8    &  50.0    &  23.8    &  35.6    &  32.7    &  53.3    & &  35.6    &  47.2    &  49.5    \\*
5737.06 &  1.06 &  $-0.81$    &  69.6    &  54.9    &  25.7    &  39.5    &  32.2    &  67.7    & &  39.7    &  55.4    &  53.0    \\*
6039.72 &  1.06 &  $-0.73$    &  70.7    &  57.0    &  33.4    &  43.7    &  49.3    &  64.0    & &  45.0    &  61.0    &  60.3    \\*
6058.14 &  1.04 &  $-1.37$    &  24.6    &  20.1    &  \nodata &  12.4    &  12.8    &  26.3    & &  14.5    &  16.8    &  15.7    \\*
6090.21 &  1.08 &  $-0.13$    &  \nodata &  96.7    &  69.7    &  90.0    &  76.3    &  107.2   & &  82.1    &  103.0   &  96.4    \\*
6111.65 &  1.04 &  $-0.81$    &  74.5    &  59.9    &  37.4    &  50.7    &  35.6    &  58.4    & &  37.5    &  58.0    &  55.8    \\*
6150.16 &  0.30 &  $-1.56$    &  102.6   &  84.5    &  40.6    &  58.8    &  58.8    &  98.7    & &  59.8    &  90.1    &  77.7    \\*
6199.20 &  0.29 &  $-1.49$    &  \nodata &  112.3   &  57.1    &  88.6    &  76.3    &  \nodata & &  84.4    &  117.7   &  106.6   \\*
6224.53 &  0.29 &  $-1.89$    &  80.5    &  64.4    &  28.9    &  43.0    &  40.6    &  70.5    & &  50.2    &  68.7    &  60.6    \\*
6233.16 &  0.28 &  $-1.94$    &  69.2    &  59.4    &  21.5    &  41.7    &  28.4    &  69.7    & &  33.6    &  57.2    &  52.2    \\*
6251.83 &  0.29 &  $-1.40$    &  127.6   &  101.6   &  57.4    &  80.4    &  76.0    &  119.9   & &  80.6    &  108.0   &  96.3    \\*
6266.31 &  0.28 &  $-2.27$    &  55.0    &  43.2    &  18.7    &  28.7    &  24.1    &  54.8    & &  29.8    &  46.9    &  37.8    \\*
6274.65 &  0.27 &  $-1.76$    &  95.6    &  79.0    &  40.9    &  53.1    &  44.5    &  83.4    & &  51.9    &  82.8    &  74.2    \\*
6285.15 &  0.28 &  $-1.63$    &  102.5   &  89.7    &  51.7    &  63.9    &  55.6    &  98.2    & &  70.8    &  91.6    &  \nodata \\*
6292.83 &  0.29 &  $-1.61$    &  103.4   &  83.3    &  44.3    &  74.1    &  60.9    &  91.3    & &  64.9    &  88.3    &  85.0    \\*
6531.42 &  1.22 &  $-0.97$    &  44.7    &  33.0    &  17.2    &  25.3    &  20.3    &  41.8    & &  23.8    &  32.4    &  32.0    \\*
\cutinhead{Mn I}
6013.51 &  3.07 &  $-0.34$    &  96.9    &  85.5    &  53.7    &  74.5    &  75.1    &  92.5    & &  62.9    &  78.3    &  75.4    \\*
6016.67 &  3.07 &  $-0.24$    &  94.7    &  101.9   &  63.5    &  73.5    &  74.2    &  88.3    & &  77.1    &  85.7    &  84.3    \\*
\cutinhead{Fe I}
6096.67 &  3.98 &  $-1.93$    &  35.8    &  31.6    &  25.2    &  29.3    &  30.3    &  33.6    & &  23.3    &  27.8    &  22.3    \\*
6127.91 &  4.14 &  $-1.40$    &  42.1    &  42.7    &  40.6    &  34.3    &  45.6    &  45.0    & &  34.5    &  37.4    &  32.7    \\*
6151.62 &  2.18 &  $-3.30$    &  110.5   &  104.9   &  82.9    &  95.9    &  89.3    &  102.8   & &  89.8    &  109.4   &  102.4   \\*
6157.73 &  4.07 &  $-1.26$    &  \nodata &  70.6    &  44.0    &  64.2    &  58.5    &  \nodata & &  57.3    &  70.0    &  \nodata \\*
6165.36 &  4.14 &  $-1.47$    &  42.4    &  35.8    &  \nodata &  43.1    &  29.5    &  42.5    & &  31.9    &  33.7    &  36.7    \\*
6187.99 &  3.94 &  $-1.72$    &  50.2    &  45.6    &  34.0    &  39.6    &  32.4    &  39.1    & &  38.3    &  43.7    &  40.6    \\*
6200.31 &  2.61 &  $-2.44$    &  \nodata &  125.6   &  108.6   &  111.5   &  112.4   &  118.4   & &  106.1   &  131.9   &  113.8   \\*
6213.43 &  2.22 &  $-2.48$    &  157.3   &  147.5   &  127.2   &  139.1   &  132.8   &  152.3   & &  132.7   &  160.4   &  160.5   \\*
6219.28 &  2.20 &  $-2.43$    &  164.7   &  149.0   &  136.0   &  147.2   &  137.0   &  161.2   & &  136.3   &  171.5   &  158.1   \\*
6226.74 &  3.88 &  $-2.22$    &  17.7    &  23.3    &  13.4    &  22.1    &  \nodata &  21.7    & &  19.6    &  21.1    &  16.7    \\*
6229.23 &  2.84 &  $-2.97$    &  69.6    &  58.0    &  56.3    &  57.0    &  50.0    &  55.9    & &  53.7    &  60.8    &  56.4    \\*
6246.32 &  3.60 &  $-0.88$    &  133.5   &  118.9   &  115.6   &  111.6   &  124.5   &  123.2   & &  114.3   &  126.5   &  124.0   \\*
6270.23 &  2.86 &  $-2.61$    &  89.1    &  88.4    &  71.6    &  89.0    &  82.7    &  90.4    & &  75.8    &  87.5    &  87.9    \\*
6297.80 &  2.22 &  $-2.74$    &  152.3   &  129.2   &  116.2   &  117.6   &  120.2   &  131.5   & &  121.5   &  156.4   &  142.2   \\*
6301.50 &  3.65 &  $-0.75$    &  130.8   &  126.0   &  117.6   &  125.3   &  126.4   &  124.6   & &  112.2   &  \nodata &  123.9   \\*
6311.51 &  2.83 &  $-3.23$    &  \nodata &  51.6    &  37.5    &  50.6    &  44.1    &  52.7    & &  39.7    &   50.6   &  34.9    \\*
6355.04 &  2.84 &  $-2.29$    &  121.2   &  116.6   &  91.4    &  112.2   &  98.5    &  117.8   & &  98.8    &  122.9   &  105.1   \\*
6380.75 &  4.19 &  $-1.38$    &  51.9    &  48.3    &  32.0    &  42.8    &  35.3    &  48.1    & &  42.5    &  47.2    &  38.8    \\*
6392.54 &  2.23 &  $-4.03$    &  43.2    &  48.2    &  34.5    &  38.0    &  33.6    &  48.6    & &  37.2    &  50.8    &  44.5    \\*
6393.60 &  2.43 &  $-1.58$    &  196.4   &  191.5   &  166.8   &  182.0   &  171.3   &  206.3   & &  187.0   &  222.5   &  183.0   \\*
6421.35 &  2.28 &  $-2.03$    &  194.2   &  182.4   &  162.3   &  174.8   &  162.9   &  182.8   & &  160.8   &  198.3   &  173.4   \\*
6430.85 &  2.18 &  $-2.01$    &  \nodata &  197.3   &  163.3   &  171.2   &  167.6   &  192.9   & &  168.2   &  219.5   &  186.6   \\*
6495.74 &  4.84 &  $-0.94$    &  16.7    &  17.8    &  16.5    &  16.0    &  20.1    &  20.5    & &  13.3    &  15.8    &  15.3    \\*
6498.94 &  0.96 &  $-4.70$    &  156.8   &  144.4   &  119.4   &  122.8   &  123.9   &  \nodata & &  122.9   &  163.3   &  136.9   \\*
\cutinhead{Fe II}
5991.38 &  3.15 &  $-3.56$    &  28.9    &  31.3    &  29.5    &  30.6    &  25.0    &  27.7    & &  21.2    &  31.2    &  23.4    \\*
6149.25 &  3.89 &  $-2.72$    &  22.0    &  23.3    &  21.0    &  21.3    &  21.6    &  20.4    & &  19.7    &  23.6    &  21.6    \\*
6247.56 &  3.89 &  $-2.33$    &  34.6    &  37.0    &  35.2    &  \nodata &  \nodata &  35.6    & &  40.0    &  40.7    &  35.3    \\*
6369.46 &  2.89 &  $-4.25$    &  12.1    &  16.0    &  13.2    &  17.3    &  15.8    &  \nodata & &  13.2    &  14.3    &  \nodata \\*
6416.93 &  3.89 &  $-2.74$    &  21.9    &  26.9    &  18.2    &  24.6    &  24.0    &  20.3    & &  18.5    &  25.1    &  18.4    \\*
6432.68 &  2.89 &  $-3.71$    &  \nodata &  35.8    &  35.1    &  33.4    &  31.7    &  29.0    & &  30.1    &  33.6    &  33.1    \\*
6456.39 &  3.90 &  $-2.08$    &  47.3    &  53.0    &  49.1    &  58.0    &  49.3    &  48.1    & &  51.9    &  57.8    &  44.6    \\*
6516.08 &  2.89 &  $-3.45$    &  55.4    &  56.5    &  45.7    &  48.2    &  45.3    &  48.4    & &  49.8    &  55.5    &  45.5    \\*
\\
\\
\cutinhead{Ni I}
6086.28 &  4.27 &  $-0.39$    &  \nodata &  18.9    &  16.8    &  36.3    &  15.7    &  17.1    & &  18.7    &  19.1    &  17.1    \\*
6108.11 &  1.68 &  $-2.44$    &  137.0   &  115.8   &  90.7    &  106.7   &  102.2   &  124.0   & &  107.7   &  133.0   &  120.5   \\*
6111.07 &  4.09 &  $-0.77$    &  13.9    &  14.1    &  \nodata &  23.0    &  18.0    &  \nodata & &  10.3    &  9.7     &  14.8    \\*
6175.36 &  4.09 &  $-0.45$    &  29.3    &  33.4    &  17.6    &  21.8    &  26.7    &  \nodata & &  25.6    &  25.8    &  22.1    \\*
6176.81 &  4.09 &  $-0.13$    &  43.8    &  42.7    &  35.0    &  33.4    &  34.4    &  50.5    & &  37.2    &  38.0    &  35.8    \\*
6586.31 &  1.95 &  $-2.73$    &  74.5    &  78.5    &  63.1    &  70.3    &  70.5    &  79.6    & &  74.4    &  92.9    &  83.7    \\*
6643.63 &  1.68 &  $-1.85$    &  169.9   &  157.5   &  138.7   &  155.9   &  145.9   &  169.1   & &  143.5   &  179.0   &  160.6   \\*
6772.31 &  3.66 &  $-0.84$    &  45.1    &  36.2    &  30.9    &  38.0    &  33.2    &  40.8    & &  39.2    &  42.0    &  43.0    \\*
7525.11 &  3.64 &  $-0.53$    &  65.7    &  67.1    &  57.6    &  63.5    &  64.6    &  72.7    & &  71.2    &  73.0    &  70.8    \\*
7555.60 &  3.85 &  {\phs}0.09 &  \nodata &  85.2    &  72.6    &  77.9    &  72.0    &  81.3    & &  75.7    &  95.3    &  82.5    \\*
7788.94 &  1.95 &  $-1.70$    &  159.0   &  152.3   &  127.8   &  134.9   &  128.9   &  146.0   & &  141.0   &  165.4   &  150.1   \\*
7797.59 &  3.90 &  $-0.14$    &  74.6    &  70.4    &  \nodata &  53.6    &  59.7    &  59.1    & &  63.9    &  79.4    &  65.1    \\*
\enddata
\end{deluxetable}


\clearpage

\begin{thebibliography}{}
\begin{small}
\bibitem[Anders \& Grevesse(1989)]{AG89} Anders, E. \& Grevesse, N. 
         1989, \gca, 53, 197
\bibitem[Angulo et al.(1999)]{NACRE} Angulo, C. et al. 1999, \nphysa, A656, 3
\bibitem[Armosky et al.(1994)]{ASLK94} Armosky, B. J., Sneden, C., Langer, G. E.
         \& Kraft, R. P. 1994, \aj, 108, 1364
\bibitem[Arp(1955)]{Arp55} Arp, H. C. 1955, \aj, 60, 317
\bibitem[Ashman et al.(1994)]{KMM} Ashman, K. M., Bird, C. M. \& Zepf,
         S. E. 1994, \aj, 108, 2348
\bibitem[Bell et al.(1979)]{BDG79} Bell, R. A., Dickens, J. A. \& Gustafsson,
         B. 1979, \apj, 229, 604
\bibitem[Bi\'{e}mont et al.(1991)]{Bie91} Bi\'{e}mont, E., Baudoux, M.,
         Kurucz, R. L., Ansbacher, W. \& Pinnington, E. H. 1991, \aap,
         249, 539
\bibitem[Blackwell et al.(1995a)]{BLS95} Blackwell, D. E., Lynas-Gray, A. E. \&
         Smith, G. 1995a, \aap, 296, 217
\bibitem[Blackwell et al.(1995b)]{BSL95} Blackwell, D. E., Smith, G. \& 
         Lynas-Gray, A. E. 1995b, \aap, 303, 575
\bibitem[Boothroyd \& Sackmann(1999)]{BS99} Boothroyd, A. I. \& Sackmann, I.-J.
         1999, \apj, 510, 232
\bibitem[Briley et al.(1989)]{BBSH89} Briley, M. M., Bell, R. A., Smith
         G. H. \& Hesser, J. E. 1989, \apj, 341, 800
\bibitem[Briley et al.(1994)]{BBHS94} Briley, M. M., Bell, R. A., Hesser,
         J. E. \& Smith, G. H. 1994, Can. J. Phys., 72, 772
\bibitem[Briley et al.(1990)]{BBHD90} Briley, M. M., Bell, R. A., Hoban, S.
         \& Dickens, R. J. 1990, \apj, 359, 307
\bibitem[Briley et al.(1999)]{BGA99} Briley, M. M., Grundahl, F. \& Andersen,
         M. I. 1999, \baas, 195, 3.03
\bibitem[Briley et al.(1991)]{BHB91} Briley, M. M., Hesser, J. E \& Bell, R. A.
         1991, \apj, 373, 482
\bibitem[Briley et al.(1997a)]{BSKL97} Briley, M. M., Smith, V. V., King, J., 
         \& Lambert, D. L. 1997a, \aj, 113, 306
\bibitem[Briley et al.(1996)]{BSSLBH96} Briley, M. M., Smith, V. V., Suntzeff,
         N. B., Lambert, D. L., Bell, R. A. \& Hesser, J. E. 1996, \nat, 383,
         604
\bibitem[Briley et al.(1997b)]{BSSBN97} Briley, M. M., Suntzeff, N. B., Smith,
         V. V., Bell, R. A., \& Norris, J. 1997b, \baas, 189, 1363
\bibitem[Cannon et al.(1998)]{CCBHS98} Cannon, R. D., Croke, B. F. W.,
         Bell, R. A., Hesser, J. E. \& Stathakis, R. A. 1998, \mnras,
         298, 601
\bibitem[Carbon(1982)]{Carbon1982} Carbon, D. F. 1982, \apjs, 49, 207
\bibitem[Carretta \& Gratton(1996)]{CG96} Carretta, E. \& Gratton, R. G. 1996,
         in Formation of the Galactic Halo...Inside and Out, eds. H. Morrison
         and A. Sarajedini (ASP: San Francisco), p.~359
\bibitem[Carretta et al.(1999)]{CGSB99} Carretta, E., Gratton, R. G., Sneden,
         C. \& Bragaglia A. 1999, in The Chemical Evolution of the Milky
         Way: Stars vs Clusters, eds. F. Matteuci and F. Giovanelli, in
         press
\bibitem[Caughlin \& Fowler(1988)]{CF88} Caughlin, G. R. \& Fowler, W. A.
         1988, Atomic Data and Nuclear Data Tables,  40, 283
\bibitem[Cavallo(1998a)]{PHD} Cavallo, R. M. 1998a, Ph. D. Thesis, University
         of Maryland
\bibitem[Cavallo(1998b)]{Cavallo98} Cavallo, R. M. 1998b, in Stellar Evolution,
         Stellar Explosions and Galactic Chemical Evolution, ed. Anthony
         Mezzacappa (Inst. of Physics:  Bristol), p.~357
\bibitem[Cavallo \& Nagar(2000)]{CN2000} Cavallo, R. M. \& Nagar, N. M. in prep.
\bibitem[Cavallo et al.(1996)]{CSB96} Cavallo, R. M., Sweigart, A. V.
         \& Bell, R. A. 1996, \apjl, 464, L79
\bibitem[Cavallo et al.(1998)]{CSB98} Cavallo, R. M., Sweigart, A. V.
         \& Bell, R. A. 1998, \apj, 492, 575
\bibitem[Chaboyer et al.(1998)]{Chaboyer98} Chaboyer, B., Demarque, P.,
         Kernan, P. J. and Krauss, L. M. 1998, \apj, 494, 96
\bibitem[Charbonnel(1994)]{Charbonnel94} Charbonnel, C. 1994, \aap, 282, 811
\bibitem[Charbonnel(1995)]{Charbonnel95} Charbonnel, C. 1995, \apjl, 453, 811
\bibitem[Charbonnel et al.(1999)]{CDW99} Charbonnel, C., Denissenkov, P. A.
         \& Weiss, A. 1999, to appear in The Galactic Halo -- From Globular
         Cluster to Field Stars (astro-ph/9909440)
\bibitem[Cohen, Frogel \& Persson(1978)]{CFP78} Cohen, J. G., Frogel,
         J. A. \& Persson, S. E. 1978, \apj, 222, 165
\bibitem[Cottrell \& Da Costa(1981)]{CD81} Cottrell, P. L. \& Da Costa, G. S.
         1981, \apj, 245, L79
\bibitem[Cudworth \& Monet(1979)]{CM79} Cudworth, K. M. \& Monet, D. G.
         1979, \aj, 84, 774
\bibitem[Denissenkov et al.(1998)]{DDNW98} Denissenkov, P. A., Da Costa, G. S.,
         Norris, J. E. \& Weiss, A. 1998, \aap, 333, 926
\bibitem[Denisenkov \& Denisenkova (1990)]{DD90} Denisenkov, P. A. \&
         Denisenkova, S. N. 1990, Sov. Astron. Lett., 16, 275
\bibitem[Denissenkov et al.(1997)]{DWW97} Denissenkov, P. A., Weiss, A. \&
         Wagenhuber, J. 1997, \aap, 320, 115
\bibitem[Denissenkov \& Weiss(1996)]{DW96} Denissenkov, P. A. \& Weiss, A.
         1996, \aap, 308, 773
\bibitem[Ferraro et al.(1997)]{Ferr97} Ferraro, F. R., Carretta, E.,
         Corsi, C. E., Fusi Pecci, F., Cacciari, C., Buonanno, R.,
         Paltrinieri, B.\& Hamilton, D. 1997, \aap, 320, 757
\bibitem[Ferraro et al.(1998)]{Ferr98} Ferraro, F. R., Paltrinieri, B.,
         Fusi-Pecci, F., Rood, R. T. \& Dorman, B. 1998, \apj, 500, 311
\bibitem[Fujimoto et al.(1999)]{FAK99} Fujimoto, M. Y., Aikawa, M. \& Kato, K.
         1999, \apj, 519, 733
\bibitem[Fusi Pecci \& Bellazini(1997)]{FPB97} Fusi Pecci, F. \& Bellazini, M.
         1997, in The Third Conference on Faint Blue Stars,
         eds. A. G. D. Philips, J. W. Liebert and R. A. Saffer (L. Davis:
         Schenectady), p.~255
\bibitem[Fusi Pecci et al.(1990)]{BUMP} Fusi Pecci, F., Ferraro, F. R., Crocker,
         D. A., Rood, R. T. \& Buonanno, R. 1990, \aap, 238, 95
\bibitem[Gallino et al.(1999)]{GBLTS99}Gallino, R., Busso, M., Lugaro, M.,
         Travaglio, C. \& Straniero, O. 1999, to appear in Proc. of
         35$^{\rm th}$ Liege International Astrophysics Colloquium, eds.
         A. Noels et al., in press
\bibitem[Gilroy \& Brown(1991)]{GB91} Gilroy, K. K. \& Brown, J. A. 1991, \apj,
         371, 578
\bibitem[Grundahl(1999)]{FG99} Grundahl, F. G. 1999, in Spectrophotometric
         Dating of Stars and Galaxies, eds. I. Hubeny, S. R. Heap and R. H.
         Cornett (ASP: San Francisco). p.~223
\bibitem[Gustafsson et al.(1975)]{MARCS} Gustafsson, B., Bell, R. A.,
         Eriksson, K. \& Nordlund, {\AA}. 1975, \aap, 42, 407
\bibitem[Hogg(1973)]{Hogg73} Hogg, H. S. 1973, Publ. David Dunlap Obs.,
         3, No. 6
\bibitem[Holweger et al.(1995)]{HKB95} Holweger, H., Kock, M. \& Bard, A.
         1995, \aap, 296, 233
\bibitem[Holweger \& M\"{u}ller(1974)]{HM74} Holweger, H. \& M\"{u}ller,
         E. A. 1974, \solphys, 39, 19
\bibitem[Iben(1967)]{Iben67} Iben, I. Jr. 1967, \apj, 147, 634
\bibitem[Ivans et al.(1999)]{Ivans99} Ivans, I. I. et al. 1999, \aj, 118, 1273
\bibitem[Johnson \& Bolte(1998)]{JB98} Johnson, J. A. \& Bolte, M. 1998, \aj,
         115, 693
\bibitem[Kostik et al.(1996)]{KSR96} Kostik, R. J., Shchukina, N. G. \&
         Rutten, R. J. 1996, \aap, 305, 325
\bibitem[King et al.(1998)]{KSB98} King, J. R., Stephens, A. \& Boesgaard, A. M.
         1998, \aj, 115, 666
\bibitem[Kraft(1994)]{Kraft94} Kraft, R. P. 1994, \pasp, 106, 553
\bibitem[Kraft et al.(1992)]{KSLP92} Kraft, R. P., Sneden, C., Langer, G. E.
         \& Prosser, C. F. 1992, \aj, 104, 645
\bibitem[Kraft et al.(1993)]{KSLS93} Kraft, R. P., Sneden, C., Langer, G. E.
         \& Shetrone, M. D. 1993, \aj, 106, 1490
\bibitem[Kraft et al.(1995)]{KSLSB95} Kraft, R. P., Sneden, C., Langer, G. E.,
         Shetrone, M. D. \& Bolte, M. 1995, \aj, 109, 2586
\bibitem[Kraft et al.(1999)]{Kraft99} Kraft, R. P., Peterson, R. C.,
         Guhathakurta, P., Sneden, C., Fulbright, J. P. \& Langer, G. E.
         1999 \apjl, 518, L53
\bibitem[Kraft et al.(1997)]{Kraft97} Kraft, R. P., Sneden, C., Smith, G. H.,
         Shetrone, M. D., Langer, G. E. \& Pilachowski, C. A. 1997, \aj, 113,
         279
\bibitem[Langer et al.(1993)]{LHS93} Langer, G. E., Hoffman, R. \& Sneden, C.
         1993, \pasp, 105, 301
\bibitem[Langer et al.(1997)]{LHZ97} Langer, G. E., Hoffman, R. D., \& Zaidins,
         C. S. 1997, \pasp, 109, 244
\bibitem[Langer et al.(1986)]{LKCF86} Langer, G. E., Kraft, R. P., Carbon, D.
         F. \ Friel, E. 1986, \pasp, 98, 473
\bibitem[Lattanzio(1999)]{Lattanzio99} Lattanzio, J. C. 1999, in Proceedings
         of Nuclei in the Cosmos V, eds. N. Orantzos and S. Harissopulos,
         (Editions Frontieres), p~163
\bibitem[Lattanzio et al.(1999)]{LFC99} Lattanzio, J., Forestini, M. \&
         Charbonnel, C. 1999, to appear in The Changes in Abundances in 
         Asymptotic Giant Branch Stars (astro-ph/9912298)
\bibitem[Lehnert et al.(1991)]{LBC91} Lehnert, M. D., Bell, R. A., \& Cohen,
         J. G. 1991, \apj, 367, 514
\bibitem[Ludendorf(1905)]{L05} Ludendorf 1905, Publ. Astrophys. Obs. Potsdam,
         15, No. 50
\bibitem[Moore et al.(1966)]{MMH66} Moore, C. E., Minnaert, M. G. J. \&
         Houtgast, J. 1966, The Solar Spectrum 2935{\AA}  to 8770{\AA},
         2nd Ed., (Washington, D.C.: National Bureau of Standards)
\bibitem[Nave et al.(1994)]{Nav94} Nave, G., Johansson, S., Learner, R. C. M.,
         Thorne, A. P. \& Brault, J. W. 1994, \apjs, 94, 221
\bibitem[Peterson(1983)]{Peterson83} Peterson, R. C. 1983, \apj, 275, 737
\bibitem[Pilachowski et al.(2000)]{PSKHW2000} Pilachowski, C. A., Sneden, C.,
         Kraft, R. P., Harmer, D. \& Willmarth, D. 2000, \aj, in press
\bibitem[Pilachowski et al.(1996)]{PSKL96} Pilachowski, C. A., Sneden, C.,
         Kraft, R. P., \& Langer, G. E. 1996, \aj, 112, 545
\bibitem[Piotto et al.(1997)]{PIOTTO97} Piotto, G. et al. 1997, in Advances
         in Stellar Evolution, ed. R. T. Rood and A. Renzini (Cambridge:
         Cambridge Univ. Press), p~84
\bibitem[Powell et al.(1999)]{Powell99} Powell, D. C., Iliadis, C.,
         Champagne, A. E., Grossman, C. A., Hale, S. E., Hansper, V. Y.
         \& McLean, L. K. 1999, \nphysa, 660, 349
\bibitem[Reimers(1975)]{Reimers75} Reimers, D. 1975, in M\'{e}moires de la
         Societ\'{e} Royale des Sciences de Li\`{e}ge, 6e serie, tome VIII,
         Probl\`{e}mes D'Hydrodynamique Stellaire, p.~369
\bibitem[Rich et al.(1997)]{RICH97} Rich, R. M. et al. 1997, \apjl, 484, L25
\bibitem[Sandage(1953)]{San53} Sandage, A. R. 1953, \aj, 58, 61
\bibitem[Sandage \& Katem(1982)]{SK82} Sandage, A. \& Katem, B. 1982, \aj, 87,
         537
\bibitem[Sarajedini et al.(1997)]{SCD97} Sarajedini, A., Chaboyer, B. \&
         Demarque, P. 1997, \pasp, 109, 1321
\bibitem[Shetrone(1996a)]{S96a} Shetrone, M. D. 1996a, \aj, 112, 1517
\bibitem[Shetrone(1996b)]{S96b} Shetrone, M. D. 1996b, \aj, 112, 2639
\bibitem[Shetrone(1997)]{S97} Shetrone, M. D. 1997, in Poster Proc of IAU
         Symp. 189:  The Interaction between Observations and Theory, ed.
         T. R. Bedding (Univ. of Sydney:  Sydney), p.~158
\bibitem[Sneden(1973)]{MOOG} Sneden, C. 1973, \apj, 184, 839
\bibitem[Sneden et al.(1991)]{SKPL91} Sneden, C., Kraft, R. P., Prosser, C. F.
         \& Langer, G. E. 1991, \aj, 102, 2001
\bibitem[Suntzeff(1981)]{NBS81} Suntzeff, N. B. 1981, \apjs, 41, 1
\bibitem[Suntzeff \& Smith(1991)]{SS91} Suntzeff, N. B. \& Smith, V. V. 1991, 
         \apj, 381, 160
\bibitem[Sweigart(1997a)]{AVS97a} Sweigart, A. V. 1997a, \apjl, 474, L23
\bibitem[Sweigart(1997b)]{AVS97b} Sweigart, A. V. 1997b, in The Third Conference
         on Faint Blue Stars, eds. A. G. D. Philips, J. W. Liebert and
         R. A. Saffer (L. Davis: Schenectady), p.~3
\bibitem[Sweigart \& Catelan(1998)]{SC98} Sweigart, A. V. \& Catelan, M. 1998,
         \apjl, 501, L63
\bibitem[Sweigart \& Mengel(1979)]{SM79} Sweigart, A. V. \& Mengel, J. G.
         1979, \apj, 229, 624
\bibitem[Taylor(1982)]{Taylor} Taylor, J. R. 1982, An Introduction to Error
         Analysis: The Study of Uncertainties in Physical Measurements, ed.
         E. D. Commins (University Science Books:  USA)
\bibitem[Th\'{e}venin(1989)]{Thev89} Th\'{e}venin, F. 1989, \aaps, 77, 137
\bibitem[Th\'{e}venin(1990)]{Thev90} Th\'{e}venin, F. 1990, \aaps, 82, 179
\bibitem[Tody(1986)]{IRAF} Tody, D. 1986, "The IRAF Data Reduction and 
         Analysis System" in Proc. SPIE Instrumentation in Astronomy VI, ed.
         D.L. Crawford, 627, 733
\bibitem[Trefzger et al.(1983)]{TCLSK83} Trefzger, C. F., Carbon, D. F., 
         Langer, G. E., Suntzeff, N. B. \& Kraft, R. P. 1983, \apj, 266, 144
\bibitem[VandenBerg(1999)]{vdB99} VandenBerg, D. A. 1999, in The Third
         Stromlo Symposium: The Galactic Halo, eds. B. K. Gibson, T. S.
         Axelrod and M. E. Putnam (San Francisco :ASP) p.~46
\bibitem[Von Rudloff et al.(1988)]{VVH88} Von Rudloff, I. R., VandenBerg, D.
         A. \& Hartwick, F. D. A. 1988, \apj, 324, 840
\bibitem[von Zeipel(1908)]{vZ08} von Zeipel, M. H. 1908, Ann. Obs. Paris, 25,
         F1
\bibitem[Wallace et al.(1993)]{WHL93} Wallace, L., Hinkle, K. \& Livingston,
         W. 1993, N. S. ). Technical Report \#93-001
\bibitem[Wallace et al.(1998)]{WHL98} Wallace, L., Hinkle, K. \& Livingston,
         W. 1998, N. S. ). Technical Report \#98-001
\bibitem[Wallerstein et al.(1987)]{WLO87} Wallerstein, G., Leep, E. M., \&
         Oke, J. B. 1987, \aj, 93, 1137
\bibitem[Zaidins \& Langer(1997)]{ZL97} Zaidins, C. S. \& Langer, G. E. 1997,
        \pasp, 109, 252
\bibitem[Zucker et al.(1996)]{ZWB96} Zucker, D., Wallerstein, G., \& Brown,
         J. A. 1996, \pasp, 108, 911
\end{small}
\end{thebibliography}
\end{document}